\documentclass[pdftex,nofootinbib,superscriptaddress,aps,prfluids,notitlepage,longbibliography]{revtex4}
\usepackage{multirow,enumerate,epsfig,graphics,amssymb,amsmath,subeqnarray,mathrsfs,color,xcolor,epstopdf}
\usepackage{natbib}
\usepackage{color,epsfig,graphics,amssymb,mathcmd,amsmath,subeqnarray,graphicx,amsthm,subfigure,mathrsfs}
\usepackage[colorlinks=true, citecolor=red, linkcolor=blue ]{hyperref}
\usepackage{mathtools,bm} 
\usepackage{rotating,mathrsfs}
\usepackage[normalem]{ulem}

\makeatletter
\def\sgn{\mathop{\operator@font sgn}}
\def\threevdots{\vbox{\baselineskip1\p@ \lineskiplimit\z@
  \kern6\p@\hbox{.}\hbox{.}\hbox{.}}}
\makeatother

\begin{document}
\newcommand{\sm}[1]{{\color{red}{#1}}}

\title{Collective dynamics and rheology of confined phoretic suspensions}

\author{T. Traverso}
\email{traverso@ladhyx.polytechnique.fr}
\affiliation{LadHyX, CNRS -- Ecole Polytechnique, Institut Polytechnique de Paris, 91120 Palaiseau, France}

\author{S. Michelin}
\email{sebastien.michelin@ladhyx.polytechnique.fr}

\affiliation{LadHyX, CNRS -- Ecole Polytechnique, Institut Polytechnique de Paris, 91120 Palaiseau, France}

\begin{abstract}
Similarly to their biological counterparts, suspensions of chemically-active autophoretic swimmers exhibit nontrivial dynamics involving self-organization processes as a result of inter-particle interactions. Using a kinetic model for a dilute suspension of autochemotactic Janus particles, we analyse the effect of a confined pressure-driven flow on these collective behaviours and the impact of chemotactic aggregation on the effective viscosity of the active fluid. Four dynamic regimes are identified when increasing the strength of the imposed pressure-driven flow, each associated with a different collective behaviour resulting from the competition of flow- and chemically-induced reorientation of the swimmers together with the constraints of confinement. Interestingly, we observe that the effect of the pusher (resp. puller) hydrodynamic signature, which is known to reduce (resp. increase) the effective viscosity of a sheared suspension, is inverted upon the emergence of autochemotactic aggregation. Our results provide new insights on the role of collective dynamics in complex environments, which are relevant to synthetic as well as biological systems. 
\end{abstract}

\maketitle

\section{Introduction} \label{sec:intro}

The dynamics of microscopic swimmers are dominated by viscous forces, and their self-propulsion can be achieved only by nonreciprocal fluid forcing~\citep{Purcell_1977}. Phoretic particles do so by means of interfacial forces that drive a thin boundary-layer flow near the surface of the particle~\citep{anderson89}. At the typical scale of the colloidal particle, this layer's thickness is negligible so that the interfacial flow appears as a net slip velocity at the fluid-solid interface~\citep{julicher2009generic}. By forcing a relative motion of the fluid with respect to the particle, this effective slip velocity induces a net drift of the colloid~\citep{anderson89,yadav2015anatomy}, as does the cilia-driven flows of many microorganisms~\citep{Brennen1977,blake_1971}.  
When interfacial forcing and drift result from local gradients of chemical concentration, it is referred to as diffusiophoresis~\citep{anderson89}, and as self-diffusiophoresis when such gradients are generated by the particle itself via surface chemical reactions. Janus colloids represent a now-canonical example of such phoretic colloids, and generate the  gradients required for propulsion through the differential coatings of their two halves resulting in an asymmetric chemical activity~\citep{Howse2007,yadav2015anatomy,moran2017}.

Self-diffusiophoretic swimmers are chemically-active and actuate the fluid around them \citep{julicher2009generic}; thus, they can interact via the chemical and hydrodynamic disturbances they induce on their environment~\citep{sen2009chemo,Theurkauff2012,campbell2019}, like many of their biological counterparts~\citep{Lushi2012,budrene1991complex,Drescher2010}. Within sufficiently large active suspensions,  
 long-range chemical 
 or hydrodynamic interactions can cause the emergence of collective dynamics~ \citep{ibele2009schooling,yadav2015anatomy,Dombrowski2004,Saintillan2008} characterised by correlated motion of the particles~\citep{Dunkel2013,Role_of_Correlations_Morozov2017}. 
Because it results from inter-particle interactions, the correlation length $l_c$ of such collective motion is intrinsic to the suspension and is typically larger than the interaction range~\citep{Balescu_short}, thus much larger than the typical size of the swimmers. 

A second important length scale within an active suspension is $l_e$ which characterises its environment, and can be the typical size of regions with different background flow conditions or the degree of confinement (e.g. gap between obstacles, width of a channel or radius of a droplet hosting the suspension). When $l_c\sim l_e$, the collective behaviour of active systems is not suppressed but  interestingly modified, as suggested by Ref.~\cite{Wioland_2016}, who showed that the turbulent-like dynamics emerging in suspensions of \emph{E. Coli} transition to collective directional motion when the system is confined within a sufficiently narrow and closed channel.  Previously, Refs.~\cite{Wioland2013} and \cite{Lushi2014} also showed how the collective motion of bacteria confined into a small droplet induced a steady single-vortex state due to the curvature of the boundaries.

The interaction of active self-propelled particles with rigid boundaries under confinement has an impact on the particles' motion even in the absence of intrinsic collective dynamics.
In this regard, much research has focused on (steric or fluid-mediated) wall-particle interactions  at the level of an individual swimmer in order to explain a variety of experimental observations. These include the attraction of swimmers toward walls and subsequent reorientation parallel to the surface~\citep{spagnolie_lauga_2012,Li2011}, their increased residence time near the surface~\citep{Drescher2011}, the orbits of rod-like autophoretic colloids around small obstacles~\citep{Takagi2014}, scattering dynamics of swimming microalgae off of circular pillars~\citep{Contino2015}, and the influence of ciliary contact interactions with surfaces for flagellated microorganisms~\citep{kantsler2013ciliary}.  

Despite the complexities arising in the detailed description of each system,
the tendency of swimming particles to spend most of their time near boundaries appears common to many active suspensions~\citep{ROTHSCHILD1963,berke2008hydrodynamic,li2009accumulation}. Interestingly, this behaviour can also be rationalized by involving only the combined effect of self-propulsion, steric exclusion by the wall and diffusive processes~\citep{Elgeti2013,ezhilan_saintillan_2015}. Thus, at large time scales compared to those characterising the ballistic run of a swimmer, its shape and the specifics of its swimming kinematics are not necessary ingredients to predict a swimmer's larger residence time near walls.

Another key feature of natural environments is the presence of an external stimulus. This could be some non-uniform flow conditions, such as those generated by muscular contractions or heat convection in biological system~\citep{riffell2007sex} or by imposed pressure gradients in microfluidic devices ~\citep{liu2020optimised}. Other stimuli include external attractive fields such as light for phototactic micro-algae \citep{Rafai2016_rapidPRE} or synthetic swimmers \citep{sen2009chemo}. In an experiment involving a confined suspension of phototactic algae, Ref.~\cite{Rafai2016}  tested the combined effect of two simultaneous stimuli: (i) a background pressure-driven shear flow and (ii) a directional source of light. The result is the focusing of the swimmers either at the centre of the channel or at the boundaries, depending on the relative directions of the flow and light source. A similar behaviour was observed by Ref.~\cite{kessler1985_focusingGyro} for gyrotactic swimmers, where the effect of light is replaced by that of gravity. More recently, Ref.~\cite{Rusconi2014} showed how the presence of an externally-imposed pressure-driven flow affects fundamental microbial processes (e.g. nutrient uptake) by hampering chemotaxis while promoting surface attachment. 

At the typical length scales of the suspension, the  dynamics and trajectory of individual microswimmers are blurred and the system can be regarded as a continuum, namely as an \emph{active fluid}. Active fluids are known to respond in a peculiar way to external stimulations. In particular, the rheology of active suspensions was analysed experimentally for elongated pusher-like~\citep{Gachelin2013,Lopez2015} and puller-like swimmers~\citep{Rafai2010}, which were found respectively to decrease and increase the effective viscosity of the fluid as a result of the energy injection at the particle scale. These predictions are qualitatively captured by theoretical models which consider the swimmers as elongated bodies and completely neglect the presence of boundaries~\citep{Hatwalne2004,Saint_2010}. More complex models have included the effect of boundaries and inter-particle interactions in one-dimensional channels, i.e., considering inhomogeneities only along the cross-stream direction and an homogeneous streamwise direction \citep{Matilla2016}. However, the effective viscosity of an active suspension undergoing self-driven collective dynamics remains largely unexplored. Doing so would require (i) to account for spatial inhomogeneities also in the streamwise direction \citep[i.e collective dynamics are minimally captured in two  dimensions, see][]{Lushi2012,Gao2017}, and (ii) to include hydrodynamic and chemical interactions, which drive the underlying self-organization processes.
 
Having identified confinement and external stimuli as building blocks to simulate a realistic environment of active suspensions~\citep{guasto2012fluid,tufenkji2007modeling}, the aim of this work is to study the collective response of autophoretic suspensions when placed into a channel pressure-driven flow, and how such response influences its macroscopic properties, e.g. the rheological properties of this active fluid.  
To this end, the kinetic model recently used by Ref.~\cite{Traverso2020} to model autophoretic suspensions in a bulk environment is adapted here to include the effect of rigid no-slip walls and of an external flow.

In this work, we focus on suspensions of chemically-active Janus spheres whose surface properties enable their reorientation along gradients of a chemical solute produced at their surface thus making them auto-chemotactic, i.e. able to rotate and self-propel toward other particles or their chemical trails. 
At sufficiently large time scales compared to that of a ballistic swimmer's run, this effective attraction is known to promote particle self-organization in the form of \emph{asters}~\citep{Saha2014}, similar to that observed for autochemotactic microorganisms performing run-and-tumble dynamics~\citep{budrene1991complex,Lushi2018}.  

Our first objective is to investigate how such chemically-driven astering process is influenced by the confined environment and particle-generated hydrodynamic field, for different intensities of the background pressure-driven flow. We then characterise the effect of such dynamics on the coherent hydrodynamic forcing exerted by the particles, 
and thus on the apparent viscosity of the active fluid. 
Finally, we derive and study the linear stability of a reduced-order model (ROM) to identify and capture the minimal physical ingredients to explain the rich collective behaviours in the numerical simulations of the complete model.    

By accounting for both confinement and a background pressure-driven flow, thus reproducing conditions that are closer to the ones observed in practice, we predict and explain new dynamical regimes of autophoretic chemotactic suspensions and their link with the microscopic features of the particles. This represents a  step forward in the design of control strategies for active suspensions, in order to accomplish medical tasks, such as drug delivery~\citep{Akhil_SM_20_Wang2013,Mostaghaci2017} or non-invasive diagnostic tests for cancer cells ~\citep{mager2006bacteria}, or to overcome environmental challenges, e.g. nuclear waste removal~\citep{Ying2019} or \emph{in situ} bioremediation~\citep{steffan1999field,Tufenkji2007porous}, and to design active fluids with controllable rheological properties.  

The manuscript is organized as follows: Sec.~\ref{sec:KinModel} introduces the kinetic model for the suspension dynamics under confinement, and the techniques employed for its numerical solution. Then, the system of equations is solved numerically for different strengths of the background flow in Sec.~\ref{sec:NumSym},  and four different regimes are identified, each displaying different dynamics emerging from the interplay between inter-particle interactions, confinement and background flow. 
Section~\ref{sec:rheo} discusses the effects of the active stresses induced by the particles on the particle-induced flows and the effective viscosity of the suspension. Section~\ref{sec:ROM} proposes a reduced-order model for the suspension's dynamics and analyses its linear stability.  
Finally, Sec.~\ref{sec:conclusions} summarizes the main conclusions of this analysis and presents further perspectives.
\section{Kinetic model of a confined phoretic suspension} \label{sec:KinModel}

\subsection{Governing equations of the suspension dynamics} \label{sec:dim_gov_eq}
\begin{figure} 
   \centerline{ 
        \includegraphics[scale=0.8]{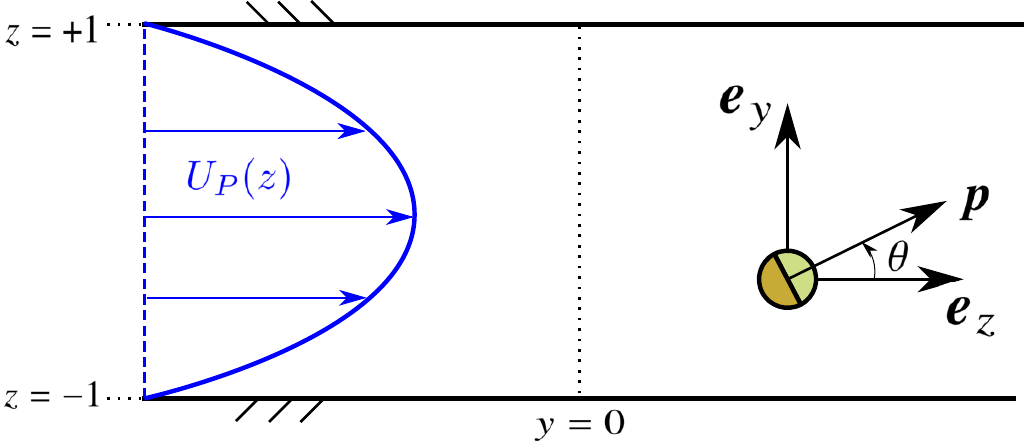}} %
    \caption{Channel geometry, imposed flow and frame of reference. } \label{fig:refframe}
\end{figure}
We analyse the dynamics of a dilute suspension of self-propelled spherical autophoretic Janus particles
confined between two parallel flat plates separated by a distance $2H$ and placed in an externally-imposed pressure-driven flow as illustrated in figure~\ref{fig:refframe}. On length scales much larger than the particle radius $R$, the probability to find a particle at a given position $\boldsymbol{x}$ with a set orientation $\boldsymbol{p}$ is described, at time $t$, by the distribution function $\Psi(\boldsymbol{x},\boldsymbol{p},t)$. Phoretic particles emit a chemical solute  and generate a net fluid slip along their surface in response to local concentration gradients, so that on their surface
\begin{equation}
D_c\boldsymbol{\hat{{r}}}\cdot\nabla C=-A(\hat{\boldsymbol{r}}),\qquad 
   \boldsymbol{u}^* = M( \hat{\boldsymbol{r}})(\boldsymbol{I}-\hat{\boldsymbol{r}} \hat{\boldsymbol{r}})  \boldsymbol{\cdot}   \nabla_x C
\end{equation}
with $C(\boldsymbol{x},t)$ the local solute concentration and $D_c$ its diffusivity, $A(\hat{\boldsymbol{r}})$ and $M(\hat{\boldsymbol{r}})$ are the (spatially-dependent) surface activity and mobility of the particle and $\hat{\boldsymbol{r}}$ the unit normal at the particle's surface. 

In a three-dimensional space $\Psi$ is a function of six independent variables, namely three spatial coordinates, two angles and time, making it unpractical for numerical time marching. In order to gain a relevant physical insight on the dynamics of the confined suspension and reduce the computational cost, we focus here on a two-dimensional case (all quantities are invariant in the $x$ direction), counting  four independent variables, namely two spatial coordinates $\boldsymbol{x}=(y,z)$, the orientation $\theta$ of the particles' director $\boldsymbol{p}=(\cos\theta,\sin\theta)$, and time $t$. Such 2D approximation is often made to  solve continuum models describing active matter numerically, and it is consistently found to provide a qualitatively accurate description of the systems' dynamics 
\citep{Saintillan2008,Lushi2018,Lushi2012,Gao2017}.   

The evolution of the suspension then follows a Smoluchowski equation (expressing the conservation of the particles in space and orientation)
\begin{align}
         \frac{\partial \Psi}{\partial t} &= -\nabla_x\boldsymbol{\cdot}(\Psi \dot{\boldsymbol{x}}) - \nabla_p\boldsymbol{\cdot}(\Psi \dot{\boldsymbol{p}}) , \label{EvolEqPsi}
\end{align}
where $\nabla_x$ denotes the spatial gradient. The operator $\nabla_p$ denotes the gradient operator with respect to the swimmers' orientation,  and its application on a scalar field $f(\boldsymbol{p})$ and a vector field $\boldsymbol{a}(\boldsymbol{p})$ amounts respectively to
\begin{align}
    \nabla_p f = \frac{\partial f}{\partial \theta}\boldsymbol{e}_{\theta} \hspace{0.3cm} \textrm{and} \hspace{0.3cm} \nabla_p\boldsymbol{\cdot} \boldsymbol{a} =  \frac{\partial}{\partial \theta} (\boldsymbol{a}\boldsymbol{\cdot} \boldsymbol{e}_{\theta}),
\end{align}
where $\boldsymbol{e}_{\theta}$ is the unit vector 
defined as ${\boldsymbol{e}_{\theta}=\partial\boldsymbol{p}/\partial\theta=(-\sin\theta,\cos\theta)}$.

 The probability fluxes $\dot{\boldsymbol{x}}$ and $\dot{\boldsymbol{p}}$  in Eq.~\eqref{EvolEqPsi} are obtained from the deterministic velocities of an individual particle of orientation $\boldsymbol{p}$ located at $\boldsymbol{x}$, in response to its own activity and to the hydrodynamic and phoretic mean fields, $\boldsymbol{u}(\boldsymbol{x},t)$ and $C(\boldsymbol{x},t)$, generated by the outer flow and other particles in its vicinity. Those read
\begin{align}
  \dot{\boldsymbol{x}} &= U_0\boldsymbol{p} + \boldsymbol{u} + \chi_t \nabla_x C - D_x \nabla_x[\ln(\Psi)] \label{xdot_dim} ,\\
 \dot{\boldsymbol{p}} &= \frac{1}{2}\boldsymbol{\omega}\times \boldsymbol{p} + \chi_r( \boldsymbol{p} \times \nabla_x C )\times \boldsymbol{p} - D_p \nabla_p[\ln(\Psi)] \label{pdot_dim},
\end{align}
where $\boldsymbol{\omega} = \nabla_x \times \boldsymbol{u}$ is the vorticity vector. The translational and rotational velocities in Eqs.~\eqref{xdot_dim} and \eqref{pdot_dim} are thus obtained by superimposing linearly (i) the intrinsic self-propulsion of the particles (there is no rotation for axisymmetric Janus particles), (ii) the hydrodynamic drift obtained from the hydrodynamic mean field using Faxen's laws, (iii) the chemical drift and rotation induced by a locally uniform gradient of concentration\footnote{Note that for a hemispheric Janus particle, the phoretic drift is purely along $\nabla C$, i.e. the component along $\boldsymbol{p}$ vanishes~\citep{Kanso2019}.}, and (iv) the particles' diffusion.

Because we consider here the specific case of spherical half-coated JPs, the physico-chemical properties of the colloids' surface, namely their mobility $M$ and activity $A$, are considered uniform on each hemisphere, respectively denoted $(A_f,M_f)$ in the front and  $(A_b,M_b)$ in the back.  For brevity, we also define $A^\pm = A_b \pm A_f$, the total activity ($+$) and activity contrast ($-$), respectively, with similar definitions for the mobility. The particle self-propulsion and drift properties can be obtained explicitly in terms of these characteristics as \citep[see Appendix~\ref{app:A} and][]{Traverso2020} 
\begin{eqnarray}
   U_0 = \frac{A^- M^+}{8 D_c}, \hspace{5mm} \chi_t = -\frac{M^+}{2}, \hspace{5mm} \chi_r = \frac{9}{16}\frac{M^-}{R}\cdot \label{velocities}
\end{eqnarray}

\bigskip

At such microscopic scales, inertia plays a negligible role, so that the hydrodynamic problem is governed by the incompressible Stokes equations for the fluid's velocity $\boldsymbol{u}$ and pressure $q$, and is forced (i) by the hydrodynamic active stresses $\mathbf{S}(\boldsymbol{x},t)$ generated collectively by the JPs, and (ii) by an imposed pressure gradient along the streamwise direction $\boldsymbol{f}_{P} = f_{y}\boldsymbol{e}_y$. As a result,
\begin{align}
  \nabla_x \boldsymbol{\cdot} \boldsymbol{u} &= 0, \label{divfreeu} \\
  -\eta \nabla_x^2\boldsymbol{u} + \nabla_x q &= \nabla_x \boldsymbol{\cdot} \mathbf{S}  + \boldsymbol{f}_{P}\label{StokesEq},
\end{align}
with $\eta$ the viscosity of the surrounding Newtonian fluid where the velocity field is subject to a no-slip boundary condition at the walls,
\begin{eqnarray}
  \boldsymbol{u} = 0 \ \ \ \textrm{at} \ \ \ z=\pm H. \label{BC_u}
\end{eqnarray}

Following Ref.~\cite{Saintillan2008}, the active stress produced by the swimmers at a given location, $\mathbf{S}(\boldsymbol{x},t)$, is obtained by  performing orientational averages of the stresslet produced by a particle oriented along $\boldsymbol{p}$, $ \hat{\mathbf{S}}(\boldsymbol{x},\boldsymbol{p},t)$, namely
\begin{eqnarray}
\mathbf{S}(\boldsymbol{x},t) =  \langle  \hat{\mathbf{S}}(\boldsymbol{x},\boldsymbol{p},t) \rangle, 
\ \ \ \textrm{where} \ \ \ \langle \bullet \rangle \equiv \int_S \bullet \Psi d\boldsymbol{p}  .          \label{TotStress}
\end{eqnarray}

As for the particle's velocities, the stresslet associated with a phoretic particle can be computed from the mobility and activity distributions on its surface and can be decomposed into two different parts $\hat{\mathbf{S}} = \hat{\mathbf{S}}_s + \hat{\mathbf{S}}_e$, namely (i) the self-induced stresslet $\hat{\mathbf{S}}_s$ corresponding to the phoretic response of the particle to its own activity, and (ii) the externally-induced stresslet $\hat{\mathbf{S}}_e$ corresponding to its phoretic response to an external solute gradient $\boldsymbol{G}$~\citep[see Appendix~\ref{app:A} and][]{Traverso2020}:
\begin{equation}
 \hat{\mathbf{S}}_e  =  \hat{\alpha}_e \left[\boldsymbol{Gp} +  \boldsymbol{pG} + (\boldsymbol{G}\boldsymbol{\cdot}\boldsymbol{p})(\boldsymbol{pp}-\mathbf{I}) \right],\qquad
 \hat{\mathbf{S}}_s  =  \hat{\alpha}_s\left(\boldsymbol{pp}-\frac{\mathbf{I}}{3}\right)  \label{stresslet_dim},
\end{equation}
 where  
\begin{eqnarray}
\hat{\alpha}_s= -\frac{10\pi\kappa\eta R^2 M^- A^-}{D_c}, \ \ \ \  \hat{\alpha}_e = \frac{15}{8}\pi \eta R^2 M^-  ,
\end{eqnarray}
with $\kappa=  0.0872$ a numerical constant. Stresslets are traceless tensors; for the present two-dimensional implementation (where $\mathbf{I}:\mathbf{I}=2$), the previous definitions must be adapted as 
\begin{equation}
 \hat{\mathbf{S}}_e  =  \hat{\alpha}_e \left[\boldsymbol{Gp} +  \boldsymbol{pG} + (\boldsymbol{G}\boldsymbol{\cdot}\boldsymbol{p})\left(\boldsymbol{pp}-\frac{3\mathbf{I}}{2}\right) \right], \qquad
 \hat{\mathbf{S}}_s  =  \hat{\alpha}_s\left(\boldsymbol{pp}-\frac{\mathbf{I}}{2}\right)  \label{stresslet_adim}.
\end{equation} 

\bigskip

At the suspension scale, the chemical concentration field $C(\boldsymbol{x},t)$ is governed by the advection-diffusion equation
\begin{align}
 \frac{\partial C}{\partial t} + \boldsymbol{u} \boldsymbol{\cdot} \nabla_xC = D_c\nabla_x^2 C - \beta_1 C + 2\pi R^2 A^+ \Phi \label{C_eq}
\end{align} 
which includes a relaxation term and a source term.  The former is proportional to $\beta_1$, which should be interpreted as a measure of the finite-time intrinsic relaxation rate of the chemical system toward its background equilibrium, i.e., in the absence of or far from all active particles. Without loss of generality, the swimmers are considered net chemical sources ($A^+>0$) and the source term is proportional to the local density of particles,
\begin{eqnarray}
\Phi(\boldsymbol{x},t) = \langle 1 \rangle . \label{Fi_def}
\end{eqnarray}

In the dilute limit, we only need to account for the dominant influence of the particles at large distances, and thus neglect the subdominant contribution of the dipolar chemical field due to the anisotropic surface activity ($A^-\neq 0$).  
The walls of the channels are chemically-inert, therefore
\begin{align}
  \frac{\partial C}{\partial z} = 0 \ \ \ \ \textrm{at} \ \ \ \ z=\pm H. \label{BC_C}
\end{align}

Finally, particles can not penetrate the channel walls so that the normal component  $\boldsymbol{e}_z\cdot\dot{\boldsymbol{x}}$ of the fluxes in Eq.~\eqref{xdot_dim} must vanish at the wall~\citep{ezhilan_saintillan_2015} providing the boundary condition on the distribution function $\Psi$
\begin{align}
 \left(U_0 \sin\theta  + \chi_t \frac{\partial  C}{\partial z}\right)\Psi = d_x\frac{\partial \Psi}{\partial z} \ \ \ \ \textrm{at} \ \ \ \ z=\pm H. \label{BC_Psi}
\end{align}
%
%

\subsection{Non-dimensional equations} \label{sec:nondim_gov_eq}
The governing equations are made dimensionless using $H$ and $H^2/D_c$, i.e. the half-width of the channel and solute diffusion time across it, as characteristic length and time, respectively. The streamwise period of the channel now reads in dimensionless form as the aspect ratio $\mathcal{A}=L/H$, and the fluid domain of interest is now defined as $-\mathcal{A}\leq y < \mathcal{A}$ and $-1\leq z\leq 1$. The reference concentration scale  $C_{\textrm{ref}} = F H A^+/D_c$ is obtained by balancing the chemical production by the phoretic particles ($nR^2A^+$) and the diffusive flux at the suspension level ($D_c C_{\textrm{ref}}/H^2$). The dimensionless parameter $F=nR^2H$ is a measure of the spatial confinement of the suspension and is obtained as the ratio of the half channel width and of the intrinsic length scale of the suspension $(nR^2)^{-1}$ introduced by Ref.~\cite{Saintillan2008}. 

Upon normalizing $\Psi$ by the conserved mean number density $n$, which is defined as
\begin{align}
 n =  \frac{1}{4LH}\int_{-H}^{H}dz\int_{-L}^{L}dy\int_S d\boldsymbol{p} \Psi(\boldsymbol{x},\boldsymbol{p},t),
\end{align}
Eq.~\eqref{EvolEqPsi} remains unchanged,  with fluxes now given in non-dimensional form by
\begin{eqnarray}
  \dot{\boldsymbol{x}} &=& \frac{u_0}{F}\boldsymbol{p} + \boldsymbol{u} + \xi_t \nabla_x C - d_x \nabla_x[\ln(\Psi)] \label{xdot} ,\\
 \dot{\boldsymbol{p}} &=& \frac{1}{2}\boldsymbol{\omega}\times \boldsymbol{p} + \frac{\xi_r}{\rho}( \boldsymbol{p} \times \nabla_x C )\times \boldsymbol{p} - d_p \nabla_p[\ln(\Psi)] \label{pdot},
\end{eqnarray}
where $\rho=R/H$ is the nondimensional particle radius and $d_x=D_x/D_c$ and $d_p=D_pH^2/D_c$ are the reduced particle diffusion coefficients. The nondimensional self-propulsion and chemically-induced drifts are obtained from the dimensional properties of the particles as
\begin{eqnarray}
 u_0 = \frac{M^+ A^- H^2nR^2}{8 D_c^2} , \hspace{4mm} 
 \xi_t = -\frac{M^+ A^+ H^2nR^2}{2 D_c^2}, \hspace{4mm} 
 \xi_r = \frac{ 9 M^- A^+ H^2nR^2}{16 D_c^2}\cdot \label{nondim_u0_xit_xir}
\end{eqnarray}

Accounting for the no-flux boundary condition  on the chemical field,  Eq.~\eqref{BC_C}, the boundary condition for the distribution function becomes
\begin{align}
\frac{u_0}{F} \sin\theta\, \Psi  = D_x\frac{\partial \Psi}{\partial z} \ \ \ \ \textrm{at} \ \ \ \ z=\pm 1. \label{BC_Psi}
\end{align}

Using $\eta D_c/H^3$ as characteristic pressure gradient, the non-dimensional Stokes equations are obtained as
\begin{align}
  \nabla_x\boldsymbol{\cdot} \boldsymbol{u} &= 0, \label{divfreeu} \\
  - \nabla_x^2\boldsymbol{u} + \nabla_x q &= \nabla_x \boldsymbol{\cdot} \mathbf{S}  + \boldsymbol{f}_{P}, \label{StokesEq_nondim}
\end{align}
with the non-dimensional stresslets defined as
\begin{eqnarray}
\alpha_s = -\pi\kappa \frac{640}{9} \frac{\xi_r u_0}{\xi_t} \ \ \ \  \textrm{and} \ \ \ \ 
\alpha_e = \frac{30}{9} \pi \xi_r F \label{alphas}.
\end{eqnarray}

The imposed nondimensional pressure gradient $\boldsymbol{f}_{P} = -\gamma_w\boldsymbol{e}_y$ produces the Poiseuille background flow given by 
\begin{align}
  U_{P}(z) = -\frac{\gamma_w}{2}(1-z^2) \label{UPo},
\end{align}
where $\gamma_w$ represents the maximum nondimensional velocity gradient at the upper wall ($z=1$) and will be used as a relative measure of the background flow intensity.

Finally, the nondimensional concentration equation becomes
\begin{eqnarray}
   \frac{\partial C}{\partial t} + \boldsymbol{u}\boldsymbol{\cdot}\nabla_x C &=& \nabla^2_x C - \beta C + 2\pi\Phi \label{C_eq_nondim}
\end{eqnarray}
where $\beta^{-1/2} = l^*/H$ is the reduced screening length $l^* = \sqrt{D_c/\beta_1}$ emerging from the finite-time relaxation of the chemical system toward its equilibrium state in the absence of particles.

\subsection{Numerical solution}
In the following, we solve the complete nonlinear dynamics of the system numerically by marching in time Eqs.~\eqref{EvolEqPsi}, \eqref{StokesEq_nondim}, and \eqref{C_eq_nondim} for the particle distribution $\Psi$, velocity field $\boldsymbol{u}$ and solute concentration $C$.  The approach followed here is pseudo-spectral and uses a Chebyshev representation in the cross-channel direction ($z$, using $N_z$ modes) and a Fourier representation in the periodic directions, $y$ and $\theta$ using $N_y$ and $N_{\theta}$ modes, respectively. 
Convergence of the results was tested by performing simulations at increasing spectral resolution (up to $N_y=N_z=128$, $N_{\theta}=32$), and the values $N_y=N_z=64$ and $N_{\theta}=32$ are chosen to perform all the simulations reported here. 

The time-dependent variables $\Psi$ and $C$ are marched in time using a semi-implicit Crank-Nicholson scheme in spectral space. For each Fourier mode in $y$ and $\theta$, the boundary conditions Eqs.~\eqref{BC_C} and \eqref{BC_Psi} couple all the Chebyshev modes in $z$. Thus, by treating diffusion terms implicitly and nonlinear terms explicitly, time marching requires the solution of $N_y N_{\theta}/4$ 1D Helmholtz equations at every time step. This is done using the Chebyshev \emph{tau}-method on a Gauss-Lobatto grid~\citep{tuckerman1989}. The Stokes equations \eqref{StokesEq_nondim} are solved using the \emph{influence-matrix} method~\citep{Kleiser1980} which ensures locally the conservation of mass to machine precision, thus avoiding sources and sinks of advected probability. 
Finally, to avoid the coupling of the $\theta$-Fourier modes with the Chebyshev modes, at a given time $t=t_n$, the value of the (nonlinear) left-hand-side in the boundary condition Eq.~\eqref{BC_Psi} is treated as a known term and guessed using a shooting method, making Eq.~\eqref{BC_Psi} a linear Neumann condition for $\Psi$. This requires to iterate the solution of the whole system of equations at each time step until convergence (typically three to five iterations per time step), an approach that was found to ensure fluctuations of the \textit{O}$(1)$ mean probability around its theoretical (conserved) value to be \textit{O}$(10^{-7})$ or less.  

The initial particle distribution $\Psi(\boldsymbol{x},\boldsymbol{p},t_0)=1/(2\pi)+\varepsilon\Psi'(\boldsymbol{x},\boldsymbol{p})$ is generated by adding small random perturbations to a uniform and isotropic distribution $\bar\Psi=1/(2\pi)$. The initial chemical concentration $C(\boldsymbol{x},t_0)$ is chosen as the purely diffusive steady state solution of Eq.~\eqref{C_eq_nondim} for the initial particle density considered, $\Phi(\boldsymbol{x},t_0)$.    

\subsection{Parameters selection}
The physical problem considered here is fully determined by fixing the nondimensional particle's radius $\rho$, the particle properties ($u_0$, $\xi_r$, $\xi_t$), the diffusion coefficients ($d_x$, $d_p$), the chemical decay rate $\beta$, and the degree of confinement to $F=1$. We focus throughout the rest of the paper on the effect of the background flow intensity ($\gamma_w$) on an auto-chemotactic suspension. The value of the other parameters are chosen as follows.

Within such suspensions particles acting as net sources of solute are effectively attracted to each other due to the combined effect of positive chemical reorientation ($\xi_r/\rho=1.25$) and self-propulsion ($u_0 = 0.5$). In addition, we consider in the following repulsive chemically-induced drift ($\xi_t=-0.5$) in order to isolate the effect of reorientation as chemical attraction. Setting $\xi_r/\rho$ and $\xi_t$ of similar magnitudes as $u_0$ results in the particles' passive drifts in a $O(1)$ chemical gradient to be of the same order as their intrinsic propulsion speed; this reflects the theoretical expectation that collective dynamics develop when the motion due to inter-particle interactions is comparable with that due to self-propulsion~\citep{Traverso2020}. 
These values correspond to dilute suspensions ($O(10^{-2})$ volume fraction) of JPs of colloidal size ($R\sim 10^{-6}$m), and small solute molecules ($D_c\sim 10^{-9}$m$^2$s$^{-1}$). These estimates further result in the stresslet intensities $\alpha_s=-0.7305$ (pusher-type swimmer) and $\alpha_e=0.3927$, see Eq.~\eqref{alphas}. For micron-sized spherical particles, typical rotational diffusion can be estimated using Einstein's relation, $D_p= k_B T/(8\pi\eta R^3)~\sim 10^{-1}$m$^2$s$^{-1}$, yielding $d_p=D_pH^2/D_c=0.25$. For self-propelled colloids, their effective translational diffusion (i.e., at large time scales compared to the duration of a ballistic run) can be estimated by $D_x=k_B T/(6\pi\eta R) + U_0^2D_p^{-1}/2 \sim 10^{-11}$m$^2$s$^{-1}$ \citep{Howse2007}, giving $d_x=D_x/D_c=0.025$. Finally, the reduced chemical decay rate is set to $\beta=\pi/2 $, which yields a screening length for the chemical decay of the same order as the channel width. This choice allows particles to interact chemically across the entire channel.

%
%
\section{Response of the suspension to the background flow intensity} \label{sec:NumSym}
\subsection{Overview of the system's dynamics} \label{sec:overview}

\begin{figure}
   \centerline{
        \includegraphics[scale=0.65]{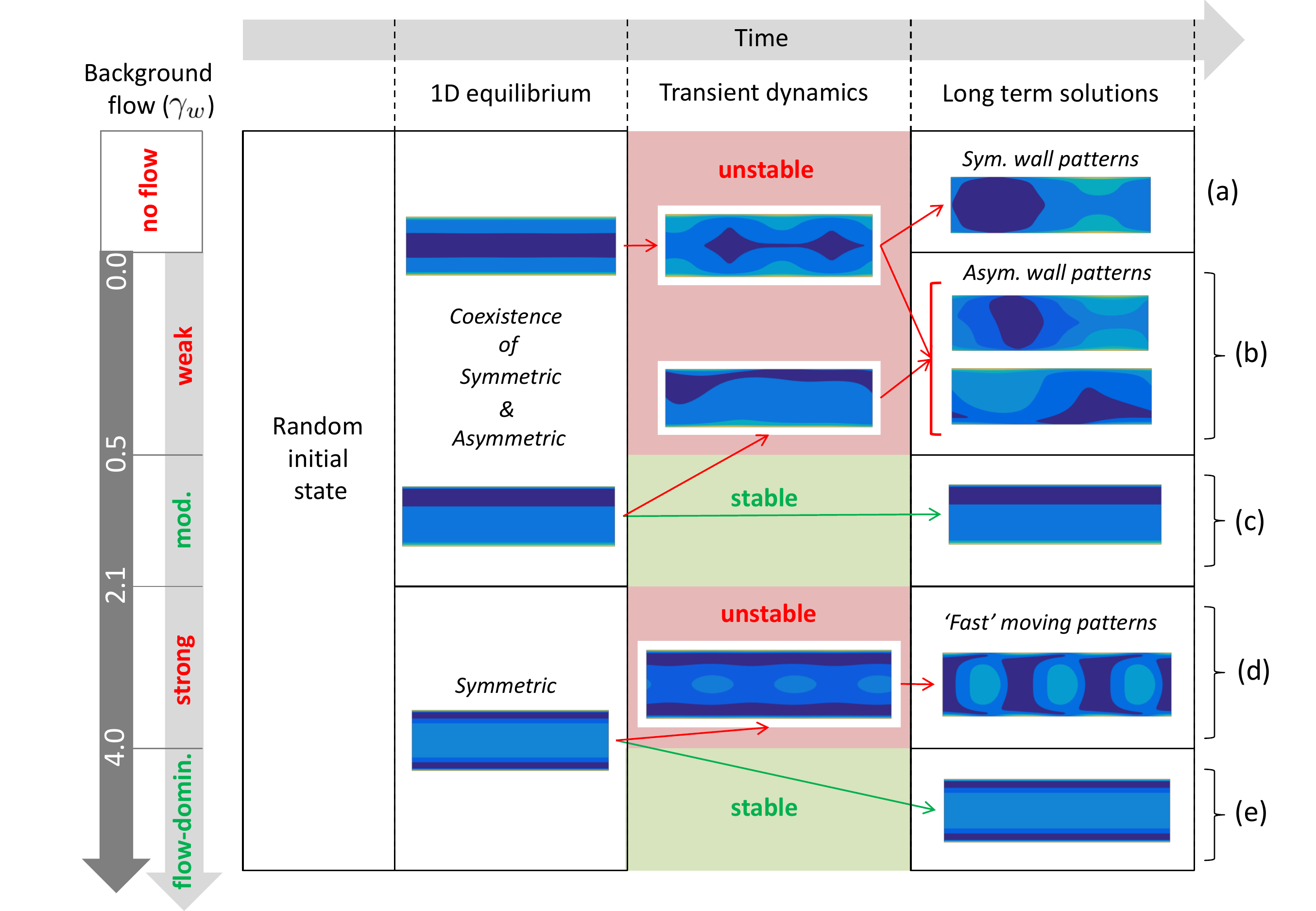}} 
    \caption{Overview of the suspension's evolution in time for increasing background flow intensity, $\gamma_w$. For each flow regime, the particle density distribution $\Phi$ is represented as the system evolves from the initial state (left) to a possibly unstable 1D fixed point (center-left). Stable (green) and unstable (red) 1D fixed points are identified and, for the latter, typical snapshots of the transient dynamics are presented (center-right), leading to the corresponding long-term solutions (right).} \label{fig:ppt_scheme}
\end{figure}

The overall dynamics of the anti-chemotactic suspension is summarized in figure~\ref{fig:ppt_scheme} for increasing background flow $\gamma_w$. For all flow intensities, starting from the initial state (a small perturbation of an isotropic suspension) and after a short transient regime, the system rapidly approaches a one-dimensional (1D) fixed point of the system, where all fields are uniform in the streamwise ($y$) direction. 
Depending on the  background flow intensity $\gamma_w$, this 1D-fixed point may however be either stable or unstable with respect to streamwise perturbations. 
Furthermore, for weak enough background flows ($\gamma_w<2.1$) two families of fixed points coexist and may be observed: one is symmetric with the channel's centre line, while the other breaks the top-down symmetry of the problem. The selection of the intermediate 1D fixed point, and therefore the initial transient dynamics, are strongly dependent on the particular choice of initial conditions -- the long-term dynamics of the system is however independent of these initial conditions and depends solely on the intensity of the background flow.
The fact that these transient regimes are indeed 1D fixed point solutions was numerically checked by solving the 1D problem (i.e. setting $\partial /\partial y = 0$ achieved in practice by computing only one Fourier mode in the streamwise direction) and then marching the system to a steady state.  
We also remark that, the long term solution of the system 
is not affected by the initial conditions and depends solely on the intensity of the background flow $\gamma_w$.

When the 1D fixed point is unstable with respect to streamwise perturbations, it appears only transiently, and, after a phase of exponential growth and saturation of the unstable modes, the solution converges toward a new $y$-dependent stable configuration, referred to in the following as long term solution. 
If, instead, the 1D equilibrium is stable with respect to streamwise perturbations, the system does not evolve away from it: the final state is stationary and uniform along the streamwise direction.

The nature of the final state depends on the background flow intensity:   we thus propose a classification of the different regimes based on the properties of the long-term solution. In the following, we characterise symmetric and asymmetric regimes based on the symmetry (or absence thereof) of the long-term solution with respect to the centreline of the channel ($y=0$). We also label these regimes as 1D or 2D, depending on whether the long-term solution is $y$-uniform (1D) or $y$-nonuniform (2D). As $\gamma_w$ is increased progressively from zero, five regimes can be distinguished based on such features as qualitatively depicted in figure~\ref{fig:ppt_scheme} (last column on the right):
\begin{enumerate}
    \item[(a)] \textbf{No-flow regime} ($\gamma_w=0$): the symmetric and asymmetric 1D fixed points are both unstable and the final state is \emph{2D and symmetric}.
    \item[(b)] \textbf{Weak-flow regime} ($0\leq\gamma_w<0.5$): the symmetric and asymmetric 1D fixed points are both unstable and the final state is \emph{2D and asymmetric}.
    \item[(c)] \textbf{Moderate flow regime} ($0.5\leq\gamma_w<2.1$): the symmetric 1D fixed point is a transient state (unstable) and the system converges at long times to the \emph{asymmetric 1D equilibrium}, which is stable.
    \item[(d)] \textbf{Strong-flow regime} ($2.1\leq\gamma_w<4$): only the symmetric 1D fixed point exists, and it is unstable; the final state is \emph{2D and symmetric}.
    \item[(e)] \textbf{Flow-dominated regime} ($\gamma_w\geq 4$): the \emph{symmetric 1D fixed point} is stable and coincides with the final state.
\end{enumerate}

In the detailed discussion of each of these regimes, the dynamics of the suspension will be shown to result from the competition of chemical interactions between particles (autochemotaxis) and background flow reorientation. We thus anticipate that in regime (a) chemical reorientation of the particle (autochemotaxis) dominates, while in regime (e) the shear-induced reorientation and flow-induced drift of the particles will be dominant. Regimes (b), (c) and (d) result from a complex interplay between these effects.   Note that the values of $\gamma_w$ defining the boundaries between regions (a), (b), (c) and (d) are obtained by performing numerical time-marching simulations and are determined numerically with a typical uncertainty $\Delta \gamma_w = \pm0.1$. 

The present study focuses specifically on the effect of shear, comparing the dynamics resulting from the entire range of shear intensity $\gamma_w$. As a result, a fixed degree of confinement is considered throughout the analysis, $F=1$. A detailed analysis of the effect of $F$ is beyond the scope of the present study; nevertheless, preliminary results (unreported here) show that reducing the degree of confinement ($F>1$) progressively weakens the wave-guide effects of the walls and is associated with a gradual transition towards bulk-like dynamics, $F\gg 1$, where particle aggregates emerge far from the walls in the form of circular asters~\citep{Traverso2020}.

%
%
\subsection{A first note on the symmetric fixed point} \label{sec:sym_base}
The initial dynamics of the suspension are characterised by a rapid relaxation toward the symmetric 1D equilibrium point; this relaxation typically occurs over a $O(\tilde{t}_0)$ time, with $\tilde{t}_0=F/u_0$ the (non-dimensional) time taken by a particle to swim across the channel. Direct relaxation toward the asymmetric fixed point requires a marked top-bottom asymmetry of the initial condition. The characteristics of this asymmetric state are further discussed in Sec.~\ref{sec:AsymState}.

Similarly to chemically-passive suspensions, and even in the absence of flow, the symmetric 1D fixed point  is characterised by a wall-normal polarisation of the swimmers induced by the impenetrability of the boundary. The corresponding boundary condition on $\Psi$, Eq.~\eqref{BC_Psi}, induces a selection in the orientation of the self-propelled particles near the walls~\citep{ezhilan_saintillan_2015}. 
These wall-normal polarisation and accumulation can be understood rather intuitively: particles pointing away from a wall will progressively swim toward the opposite side of the channel, while particles oriented toward the boundary will remain trapped for a time proportional to the characteristic time scale of rotational diffusion. 

\begin{figure}
   \centerline{
        \includegraphics[scale=0.5]{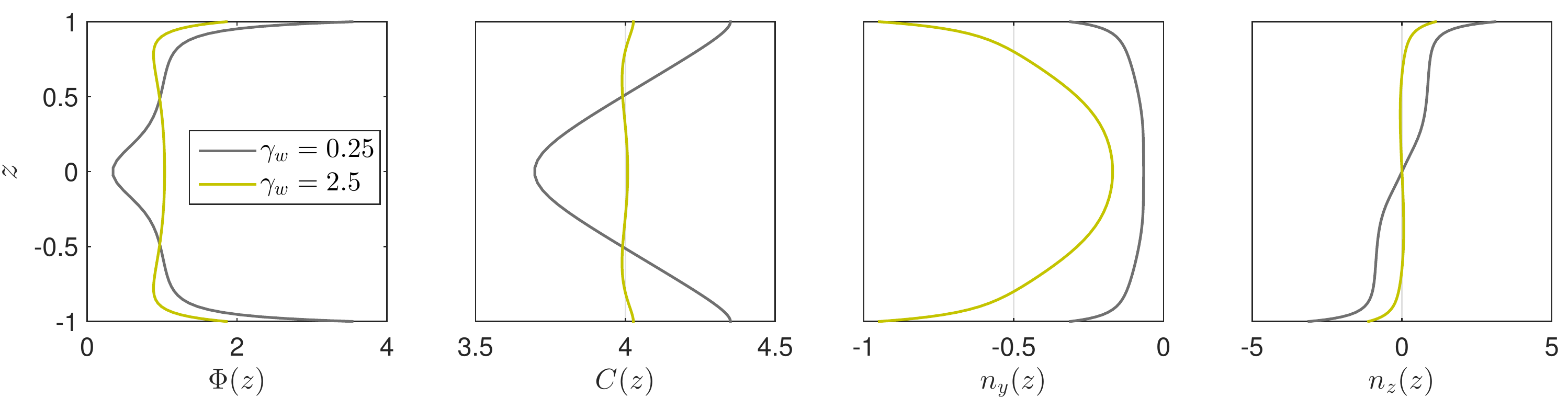}} 
    \caption{1D fixed point solution (i.e. uniform in the streamwise direction) obtained at early times ($t=40$) for weak ($\gamma_w=0.25$) and strong ($\gamma_w=2.5$) imposed flow.} \label{fig:sym_base_states}
\end{figure}

In the absence of any flow and for a fixed channel width, the thickness of the resulting polarisation/accumulation boundary layer is inversely proportional to the self-propulsion velocity, and proportional to the swimmers' rotational and translational diffusion  \citep{ezhilan_saintillan_2015}. When a background shear flow is imposed, the local vorticity induces a rotation of the particles (Faxen's law) which is  largest near the walls. In Poiseuille flow, this rotation results in upstream swimming and reduces the component of swimming toward the walls and consequent wall accumulation. These effects can be seen in figure~\ref{fig:sym_base_states} by noting $n_y<0$ and comparing the peaks of $\Phi$ and $n_z$ at $z=\pm 1$ for increasing $\gamma_w$, respectively, where $\boldsymbol{n}$ is the local polarisation defined as 
\begin{eqnarray}
   \boldsymbol{n}(\boldsymbol{x},t) = (n_y,n_z) = \langle \boldsymbol{p} \rangle \label{n_def},
\end{eqnarray} 
whose direction indicates the local expected orientation of the particles.

For weak and moderate flows ($\gamma_w < 2.1$), however, the wall accumulation remains significant and consequently, the chemical concentration generated by the particles has a marked V-shape (figure~\ref{fig:sym_base_states}). The associated chemical gradient induces the reorientation of the chemotactic  JPs ($\xi_r>0$) towards the upper (resp. lower) wall in the upper (resp. lower) half of the channel, i.e. $n_z>0$ (resp. $n_z<0$) even far from the boundaries, where the effect of confinement is still not markedly perceived. This results in the depletion of particles around the channel's axis (Figs.~\ref{fig:sym_base_states} and \ref{fig:Fi_C_NoFlow}, top). It should be noted that this centerline depletion  and reinforced wall accumulation here have a chemotactic origin and therefore differ from the shear-trapping mechanism observed for elongated swimmers \citep{ezhilan_saintillan_2015,Subramanian_2020_JFM,Rusconi2014,bearon_hazel_2015}.
%
\subsection{No imposed flow: symmetric wall particles' aggregates} \label{sec:No_flow}
In order to better understand the effect of the background flow on the suspension, we first analyse here its dynamics in the absence of flow ($\gamma_w=0$), a regime where both symmetric and asymmetric 1D equilibrium states exist but are unstable, and are thus only observed transiently (figure~\ref{fig:ppt_scheme}a). 

Autochemotactic suspensions are characterised by swimmers that modify their chemical environment and reorient in response to the perturbations produced by others. If the orientation bias is in the direction of the source of the perturbation, i.e., another swimmer, then aggregates of swimmers can form. This type of collective behaviour is observed in suspensions of living microorganisms that react to chemical cues secreted by their counterparts by modifying their tumbling rate, resulting in their biased orientation at the time scale of the collective dynamics \citep{budrene1991complex,Lushi2012}. Exploiting their front-back mobility contrast ($M^->0$), the autophoretic JPs considered here reorient along the chemical gradient generated by other particles, which act as chemical sources ($A^+>0$) \citep{Traverso2020,Liebchen2015}, see Eq.~\eqref{nondim_u0_xit_xir}, resulting in similar collective dynamics. 

\begin{figure} 
   \centerline{ 
        \includegraphics[scale=0.50]{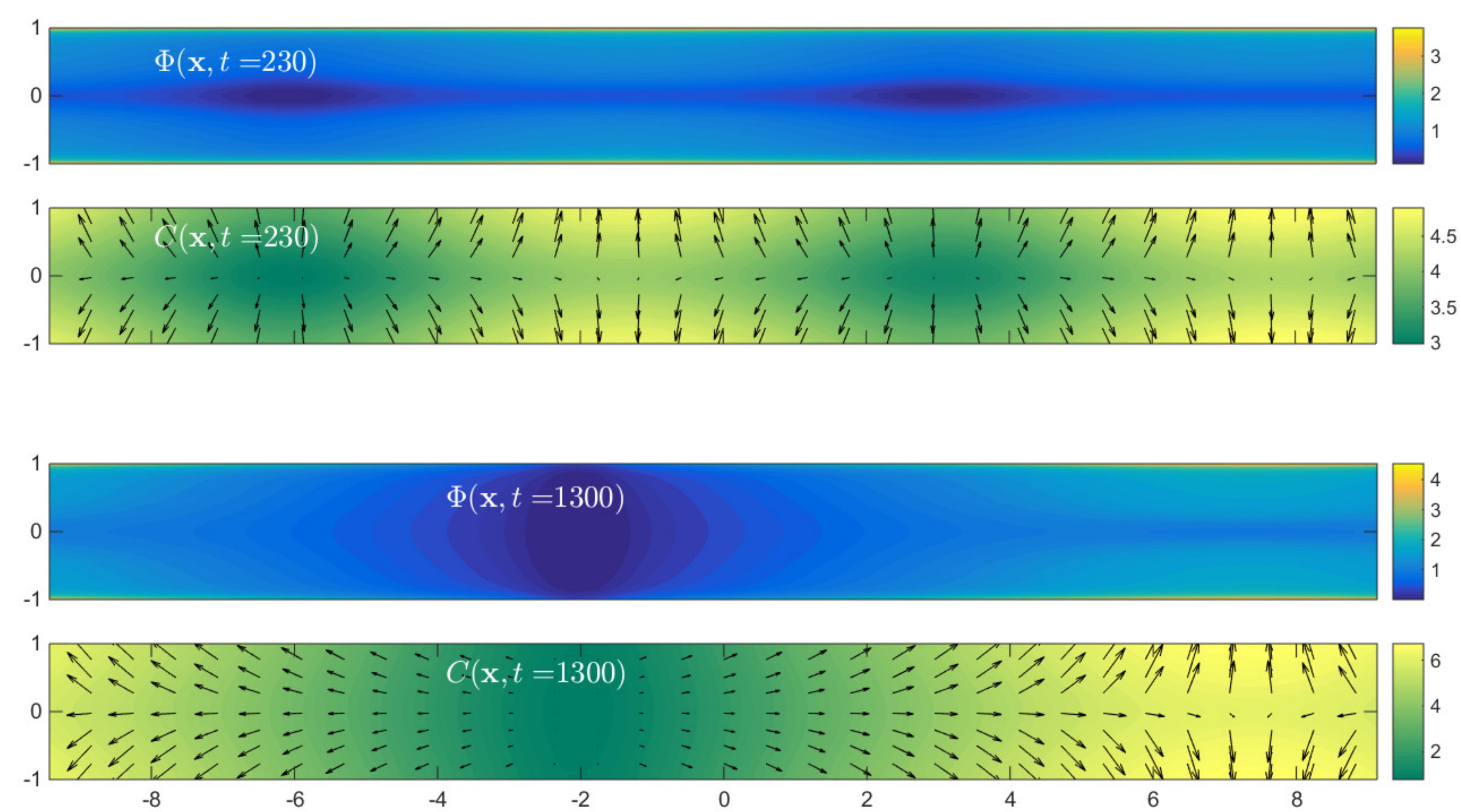}} %
    \caption{ No flow regime $\gamma_w=0$: (top, $t=230$) initial linear growth of the 2D unstable modes from the unstable 1D equilibrium; (bottom, $t=1300$) final state. In each case, the particle density $\Phi$ is reported in the top panel, and the chemical concentration $C$ together with the polarisation vector $\boldsymbol{n}$ (black arrows) in the bottom panel.
    } \label{fig:Fi_C_NoFlow}
\end{figure}

Figure~\ref{fig:Fi_C_NoFlow} reports the particle density $\Phi(\boldsymbol{x},t)$ and chemical concentration $C(\boldsymbol{x},t)$ at the onset of the instability of the symmetric 1D fixed point (top) and at large times (bottom) when no background flow is present ($\gamma_w=0$). The $y$-uniform boundary layer along the walls starts to self-organize into aggregates of particles separated by relative depletion regions. This self-organization process stems from the autochemotactic nature of the swimmers ($\xi_r>0$) which communicate chemically in the streamwise direction; this is already witnessed at the onset of the instability in the alignment of $\boldsymbol{n}$ with the local chemical field $C$, figure~\ref{fig:Fi_C_NoFlow}. 

At large times the aggregation process saturates and aggregates of particles located near the same wall merge. The final state in figure~\ref{fig:Fi_C_NoFlow} (bottom) results from the balance of (i) autochemotactic fluxes, (ii) wall-normal polarisation and accumulation, (iii)  phoretic repulsion ($\xi_t<0$) and (iv) translational/rotational diffusion of the particles.  

We finally note the left-right symmetry of the final state, as there is no special direction along the $y$-axis in the absence of imposed flow.
\subsection{Weak imposed flow: asymmetric wall particles' aggregates} \label{sec:weak_flow}

To characterise the response of the particles to a weak imposed flow, we focus in this section on the suspension dynamics for $\gamma_w=0.25$. Here again, both symmetric and asymmetric 1D fixed points are unstable and thus only observed transiently (figure~\ref{fig:ppt_scheme}b). 

\begin{figure} 
   \centerline{ 
        \includegraphics[scale=0.5]{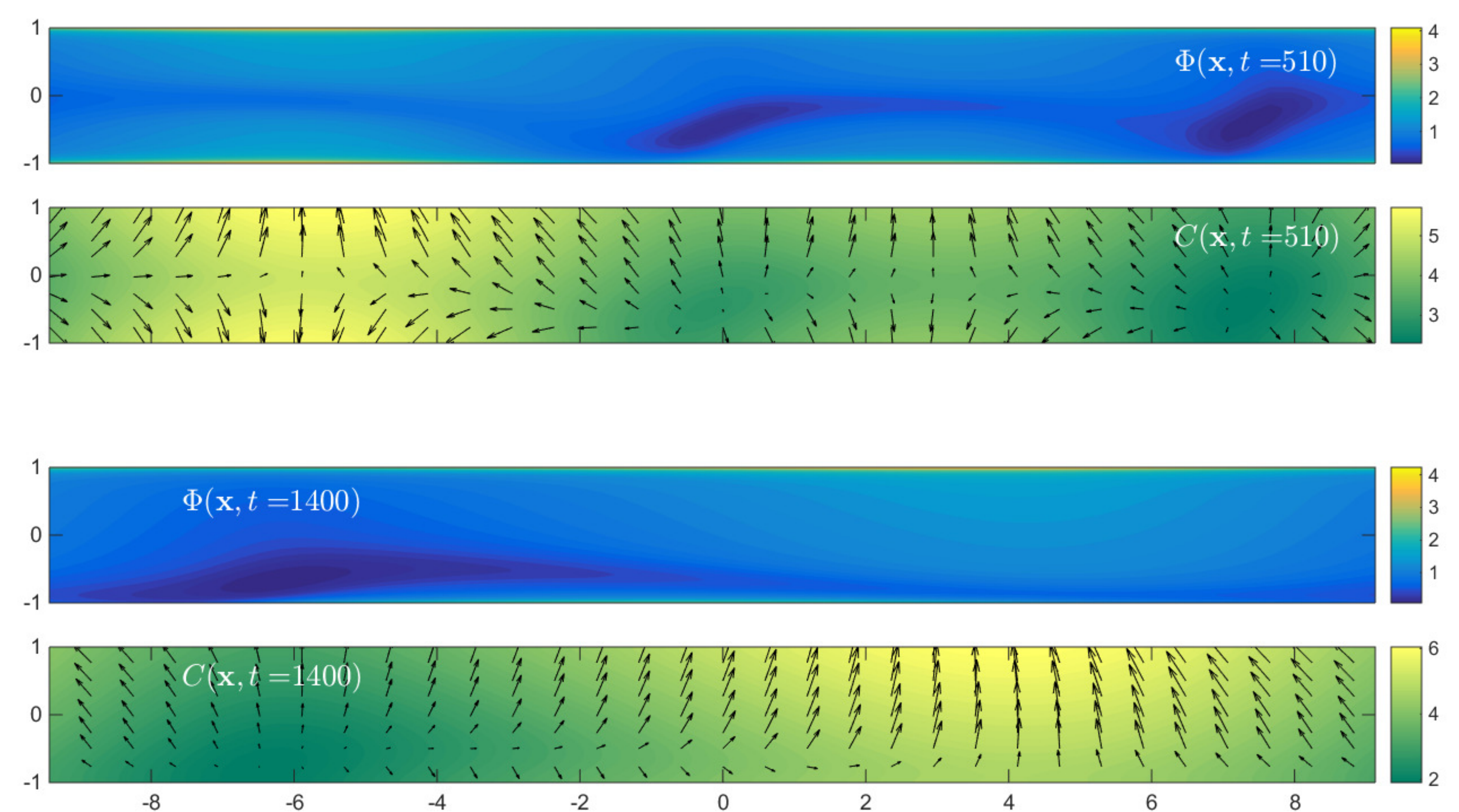}} %
    \caption{
 Weak-flow regime $\gamma_w=0.25$: (top, $t=510$) initial linear growth of the 2D unstable modes from the unstable 1D equilibrium and simultaneous top-down symmetry breaking; (bottom, $t=1400$) final state. In each case, the particle density $\Phi$ is reported in the top panel, and the chemical concentration $C$ together with the polarisation vector $\boldsymbol{n}$ (black arrows) in the bottom panel.    
   } \label{fig:Fi_C_lowSh}
\end{figure}

At early times, the transient dynamics are very similar to those found in the no-flow case, and the system's behaviour is nearly indistinguishable from that observed in figure~\ref{fig:Fi_C_NoFlow} (top), a sign that for weak flow, the effect of chemical reorientation dominates the shear-induced rotation even in the high shear region near the walls. 
Yet, as discussed in Sec.\ref{sec:sym_base}, the presence of the background flow field reduces the wall normal polarisation and accumulation. This destabilises the symmetric solutions and, at later times ($t=510$), we observe the onset of a symmetry breaking instability (figure~\ref{fig:Fi_C_lowSh}, top). At the same time aggregates located on the same side of the channel begin to merge, as in the case with no flow. The final solution is characterised by an asymmetric 2D state (figure~\ref{fig:Fi_C_lowSh}, bottom), which represents an intermediate configuration between the wall aggregates observed for no-flow, and the $y$-uniform asymmetric state which will be observed at higher $\gamma_w$ and discussed in Sec.~\ref{sec:AsymState}.    

The flow-induced left-right asymmetry of the particles' aggregates is clearly visible in figure~\ref{fig:Fi_C_lowSh}. When no flow is imposed, the horizontal polarisation, $n_y$, is perfectly antisymmetric with respect to a vertical axis cutting through the centre of an aggregate (see figure~\ref{fig:Fi_C_NoFlow}). The $y$-uniform background vorticity breaks this antisymmetry in $n_y$, and thus the left-right symmetry in $\Phi$ and $C$. 

Finally, in contrast with $\gamma_w=0$, the particles are also transported downstream by the background flow and, at large times, the solution is observed to be steady in a moving reference frame, and thus represents a travelling wave. The corresponding wave speed, which increases with $\gamma_w$, is neither the maximum nor the average flow speed but instead depends on the particle distribution within the channel. 

%
%
%
%
%
\subsection{Moderate imposed flow: asymmetric and stable 1D fixed point} \label{sec:AsymState}
\begin{figure}
   \centerline{ 
        \includegraphics[scale=0.5]{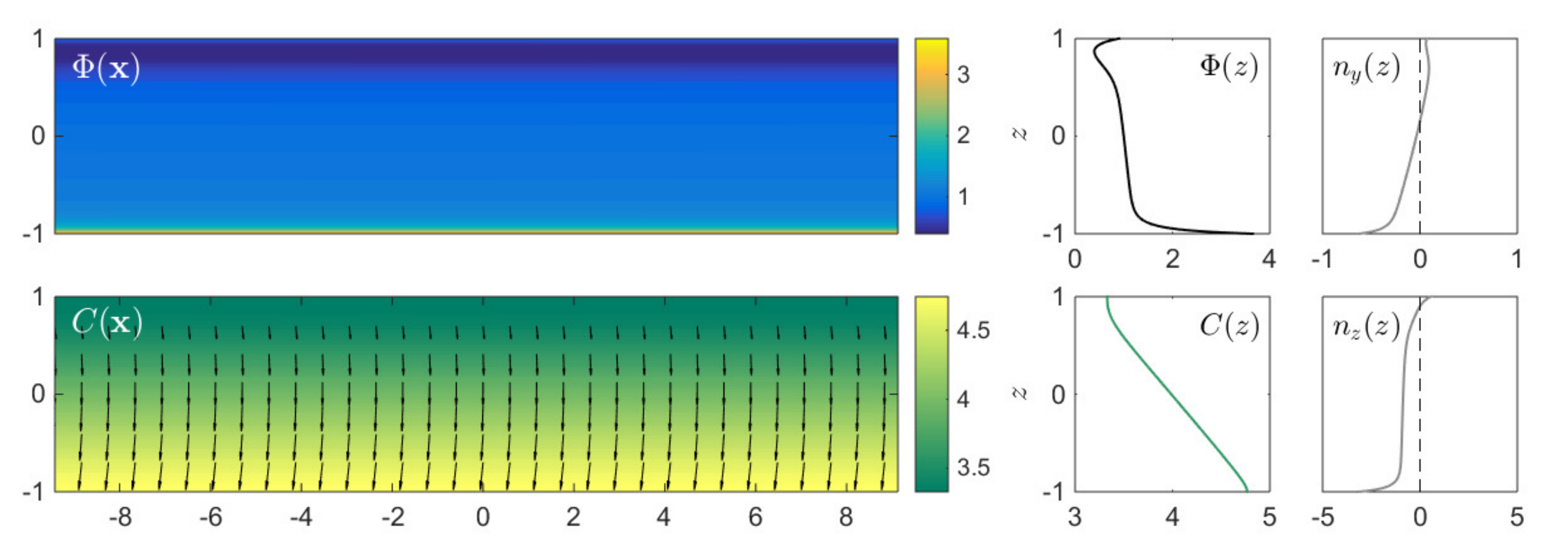}}
    \caption{Moderate shear regime ($\gamma_w=1$), long term $y$-uniform and $y$-asymmetric stable solution.} \label{fig:Fi_C_asymm}
\end{figure}
We now set $\gamma_w=1$ to analyse the moderate flow regime (figure~\ref{fig:ppt_scheme}c).  The solution first converges transiently to the (unstable) symmetric 1D fixed point as for weaker flows. In contrast with the previous regime, the most unstable mode is not two-dimensional but is instead asymmetric and uniform in the $y$-direction. At large times the solution thus converges to the stable asymmetric 1D equilibrium (figure~\ref{fig:Fi_C_asymm}). Two boundary layers near the walls retain the features of the symmetric state (Sec.~\ref{sec:sym_base}), namely a wall-normal polarisation ($n_z(-1)<0$ and $n_z(1)>0$) and a local increase in the particle density. 

In this top-down symmetry-breaking instability, a small perturbation of the symmetric state leads to an increase of chemical production by the JPs on one side of the channel. Chemotactic swimmers then reorient in response to the chemical gradient by polarizing along the $z$-axis in the direction of the high concentration side thus amplifying the initial perturbation. The final asymmetric steady state is the result of the balance between the flux due to self-propulsion, which can be visualized through the polarisation field pointing downward across most of the channel, and the upward flux due to phoretic repulsion ($\xi_t<0$) induced by the chemical gradient (see $n_z$ and $C$ in figure~\ref{fig:Fi_C_asymm}). 

This mechanism does not involve the background flow and, consistently, the 1D asymmetric equilibrium is found also for $\gamma_w=0$. The role of the flow, however, is to hinder the formation of wall aggregates thus stabilising the 1D asymmetric state. Physically, aggregates formation is triggered by chemical reorientation which can not act fast enough in comparison with vorticity-induced rotation near the walls, resulting in a misalignment of the polarisation vector with respect to the local chemical gradient.

%
\subsection{Strong imposed flow: \emph{fast} moving patterns} \label{sec:moving_patterns}
\begin{figure} 
   \centering{
        \includegraphics[scale=0.7]{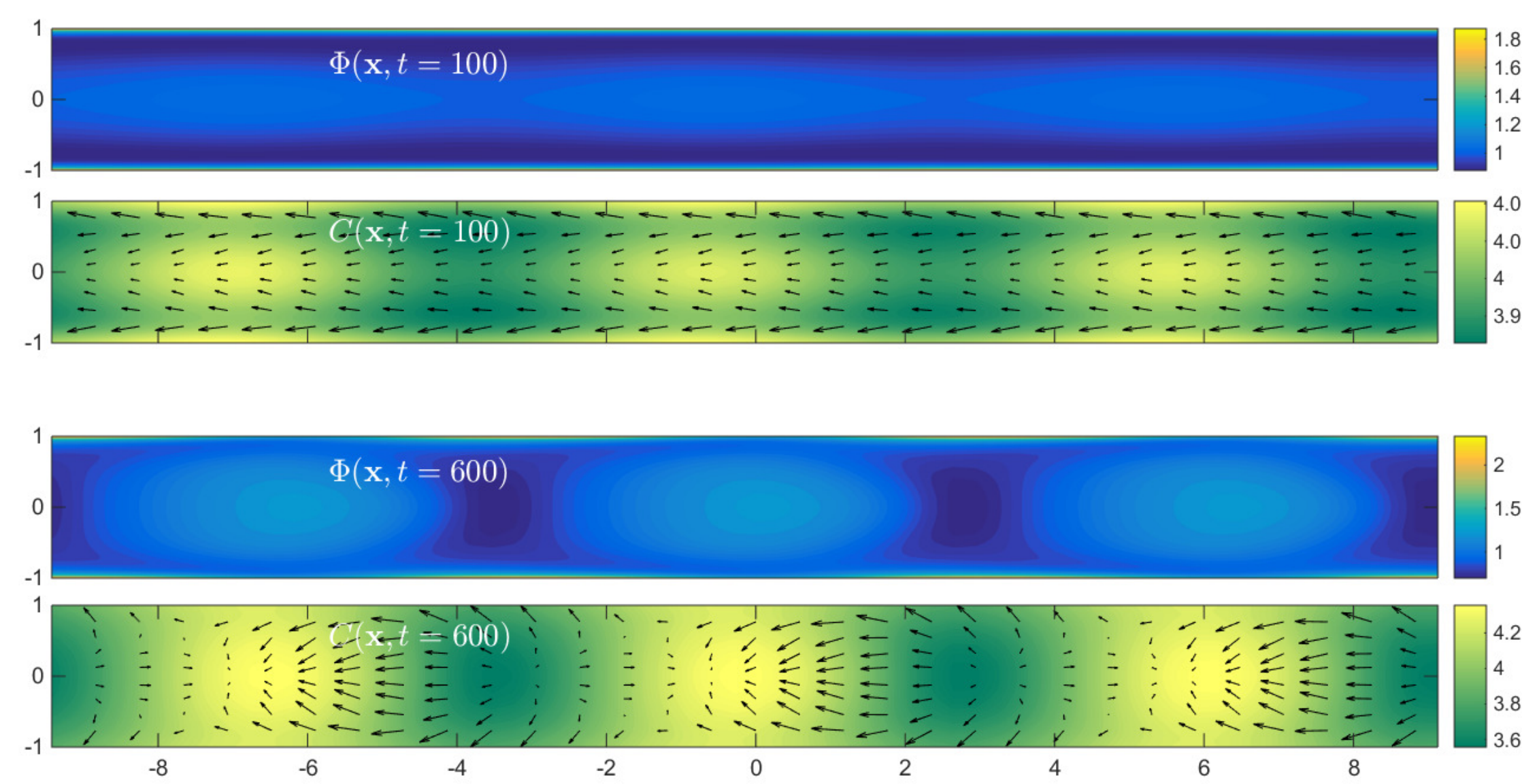}} %
    \caption{
Strong-flow regime $\gamma_w=2.5$: (top, $t=100$) initial linear growth of the 2D unstable modes from the unstable 1D equilibrium; (bottom, $t=600$) final state. In each case, the particle density $\Phi$ is reported in the top panel, and the chemical concentration $C$ together with the polarisation vector $\boldsymbol{n}$ (black arrows) in the bottom panel.     
 } \label{fig:Fi_C_highSh}
\end{figure}
\begin{figure}
   \centerline{
        \includegraphics[scale=0.41]{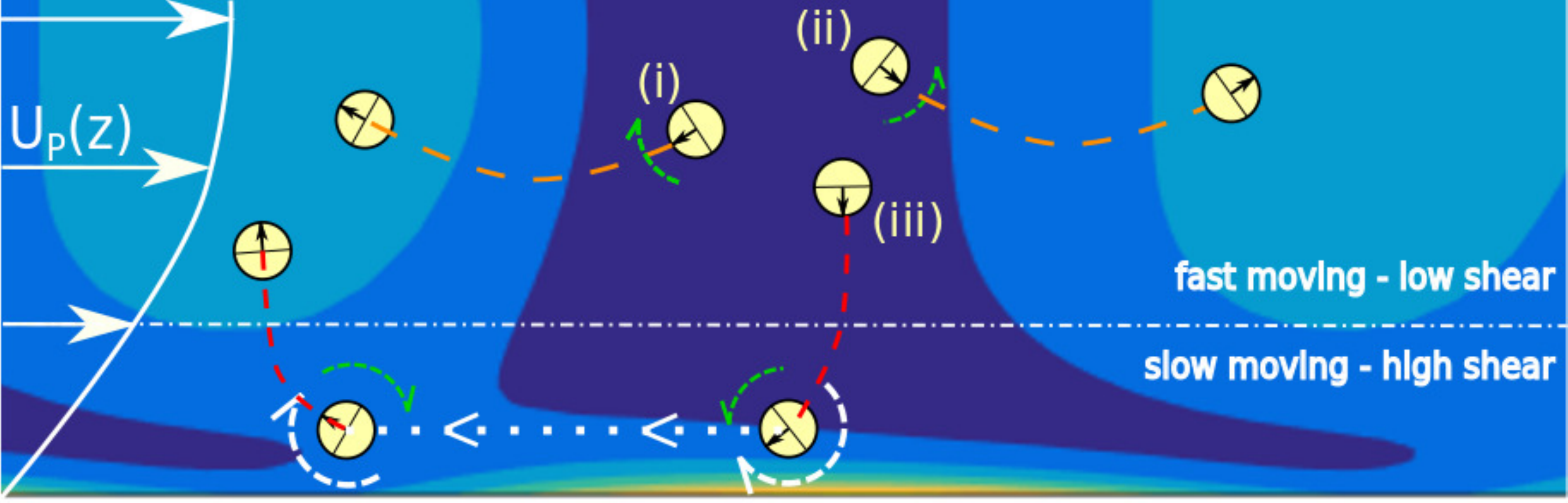}} 
    \caption{Schematics of the mechanism leading to moving patterns in the strong flow regime (see figure~\ref{fig:Fi_C_highSh}, only the bottom half of the channel is depicted here).} \label{fig:schematic}
\end{figure}
The strong-flow regime is observed for flow intensities in the range $2.1\leq\gamma_w<4$ (figure~\ref{fig:ppt_scheme}d). In that case, the 1D fixed point is characterised by a higher particle density near the centreline (figure~\ref{fig:Fi_C_highSh}): such particle trapping in the low-shear region of the channel was already observed in suspensions of nearly-spherical swimmers \citep{Barry_2015_exp,Rusconi2014}. 
From this unstable fixed point, particles  self-organize in regularly-spaced aggregates around the centre of the channel, in a process analogous to the one experienced in the bulk \citep{Traverso2020}, although it is more pronounced here in the centre of the channel, where the local background vorticity vanishes and therefore does not interfere with the chemical reorientation of the particles promoting such chemotactic instability.

Eventually the solution evolves toward a travelling wave characterised by alternating aggregation and depletion regions organised in a checkerboard pattern around the centreline and at the walls (see the particle distribution at $t=600$ in figure~\ref{fig:Fi_C_highSh}): at a given location along the channel, depletion regions at the centreline correspond to aggregation at the walls, and vice versa. Such pattern moves at a speed close to the maximum imposed velocity of the background flow, ${U_P(z=0)}$, indicating the driving of such moving patterns by the particle and solute organisation in the central low-shear region. This is in contrast with the travelling wave observed at lower shear (Sec.\ref{sec:weak_flow}), where the dynamics are dominated by the particles' aggregates near the walls and the travelling wave moves at a much lower speed compared to the fluid velocity at the centreline. 

The self-sustained existence of the moving patterns in figure~\ref{fig:Fi_C_highSh} requires a mechanism that continuously drives particles away from depletion regions and towards accumulation regions at the centreline. Such mechanism can be explained in terms of the combined effects of autochemotaxis (chemical reorientation, $\xi_r>0$), self-propulsion, background vorticity (hydrodynamic rotation) and confinement; its saturation is reached due to the effect of particle diffusion and phoretic repulsion ($\xi_t<0$). 
These dynamics are better understood by considering the individual trajectories of particles leaving a centre-line depletion region along one of the three following directions: (i) anti-parallel and (ii) parallel to the flow direction and (iii) towards the wall (figure~\ref{fig:schematic}).

Given the small local vorticity, particles (i) and (ii) are able reorient in the local chemical gradient (green circular arrows). Their trajectories remain in the fast-moving low-shear region leading towards the aggregation regions at the back and front of the depletion regions, respectively. 
Particle (iii) initially swims downward towards the slow-moving high-shear region, where the vorticity increases and the intensity of the flow gets weaker as $U_P(\pm 1)=0$. The orientation of the particle here is determined by the hydrodynamic clockwise rotation (white circular arrows) which is balanced by the chemically-induced rotation (green circular arrow). This mechanism, together with the fact that particles can not penetrate the wall, produces the high-density regions near the walls. 
Particle (iii) is now in the slow-moving region near the wall and, from the point of view of the  moving patterns near the centreline, thus travels upstream (white dotted arrow). This relative motion brings particle (iii) in a position where the chemical reorientation induced by the centreline pattern and the background vorticity both act in the same direction, directing the particle back upward toward the moving high concentration region.

\section{Effective rheology of the phoretic suspension} \label{sec:rheo}
%

\subsection{Particle-induced flow and effective viscosity}
A remarkable feature of the analysis presented in the previous section is that understanding the characteristics of the suspension dynamics didn't involve the flow forcing exerted by the particles. This suggests that hydrodynamic interactions between particles are subdominant with respect to the leading order effect of the vorticity-induced reorientation by the imposed flow or the chemical alignment of the particles with the local solute gradients. 

Yet, during their swimming motion, phoretic particles exert stresses on the surrounding fluid, and generate a modified flow field within the suspension that can alter the overall fluid transport through the channel for a fixed forcing pressure gradient. In doing so, the phoretic particles modify the effective rheology of the active suspension. More specifically, using the linearity of Stokes equations, the volumetric flow rates associated with the background and particle-induced flows are defined as 
\begin{align}
\dot{Q}_P=\int_{-1}^1U_P(z)dz, \hspace{3mm} 
\dot{Q}_d(t)= \boldsymbol{e}_y\cdot\int_{-1}^1\boldsymbol{u}_d(y^*,z,t)dz \label{eq:flowrate}.
\end{align}
The definition of $\dot{Q}_d$ does not depend on the choice of $-\mathcal{A}\leq y^*\leq\mathcal{A}$ 	due to the incompressibility of the fluid. 
The velocity ${\boldsymbol{u}_d=\boldsymbol{u}-U_P(z)\boldsymbol{e}_y}$ is the particle-induced flow field, i.e. the flow velocity generated by the particles' forcing $\mathbf{S}$ in Eq.~\eqref{StokesEq_nondim}. 

In the absence of particles, the classical Poiseuille law establishes that the driving pressure gradient is proportional to $\eta\dot{Q}_P$. Here, the pressure gradient driving the flow is fixed, so that the effective (modified) viscosity of the suspension is such that $\eta\dot{Q}_P=\eta_\textrm{eff}(\dot{Q}_P+\dot{Q}_d)$. In the following, we will characterise the rheology of the suspension through the evolution of the relative viscosity
\begin{eqnarray}
\eta_r =\frac{\eta_\textrm{eff}}{\eta}= \frac{1}{1+\dot{Q}_d/\dot{Q}_P}\cdot \label{eta_r_def}
\end{eqnarray}
For a fixed background pressure forcing, the effective viscosity of the suspension will be smaller (larger) than the solvent's if the particle-induced flow reinforces (resp. opposes) the imposed flow rate. To understand and characterise this phenomenon, we now specifically analyse the particle-induced flow field ${\boldsymbol{u}_d}$ in the different regimes identified in the previous section.

As detailed in appendix \ref{app:A} and in Ref.~\cite{Traverso2020}, the particle-induced stress on the fluid has two origins: a self-induced contribution ($\mathbf{S}_s$) corresponding to the slip distribution on the particle surface resulting from the concentration field generated by the particle itself (i.e., its stresslet if it was isolated chemically and hydrodynamically), and a chemically-induced contribution corresponding to the slip induced on a particle surface by the chemical gradients generated by all the others ($\mathbf{S}_e$). Accordingly, following the decomposition ${\mathbf{S}=\mathbf{S}_s+\mathbf{S}_e}$, and exploiting the linearity of Stokes equations, the resulting particle-induced flow field and associated volumetric flow rate can also be split into two parts 
\begin{eqnarray}
  \boldsymbol{u}_d(\boldsymbol{x},t) = \boldsymbol{u}_s + \boldsymbol{u}_e,\qquad \dot{Q}_d=\dot{Q}_s+\dot{Q}_e \label{u_decomposition}  .
\end{eqnarray}

Before proceeding by examining the \emph{global} relative viscosity $\eta_r$, we remark that other approaches could be used. One is to look at \emph{local} quantities by defining a particles' shear viscosity $\eta_p(\boldsymbol{x})$ which is the constant of proportionality between the active stresses and the strain rate of the fluid at a given location. In both approaches one needs the distribution function to be at least anisotropic for the model to predict any effect at all, i.e., to avoid the average active stresses to cancel out. Anisotropy of $\Psi$ is not sufficient and, depending on the form of $\mathbf{S}$, certain orientational moments needs be nonzero, e.g., $\langle \boldsymbol{pp} \rangle$ for  $\mathbf{S}_s$ and $\langle \boldsymbol{p} \rangle$ for $\mathbf{S}_e$.

In suspensions of elongated microorganisms, for example, a nonzero average active stress can result from the preferred orientation of the swimmer in an imposed irrotational flow with constant rate-of-strain tensor $\mathbf{E} = (\nabla\boldsymbol{u} + \nabla\boldsymbol{u}^{\textrm{T}})/2$~\citep{Saint_2010}. 
In contrast, no shear alignment occurs for spherical particles, therefore an external flow field is not sufficient to induce an anisotropic distribution. Here confinement introduces a first wall-normal preferential direction while the left-right symmetry is lost by imposing the background (rotational) flow and, finally, a nonzero active contribution to the total flow rate is predicted. As we will see, the $y$-component of the chemical gradient, which emerges when the solution is not uniform in the streamwise direction, locally introduces a new preferential direction with interesting consequences on the flow generated by the particles. 

In the following, we analyse for each of the identified flow regimes in figure~\ref{fig:ppt_scheme} the net flow rate generated by the particles. In the no-flow regime, the suspension's characteristics are left-right symmetric (there is no imposed flow direction breaking the symmetry): the net particle-induced volumetric flow rate is therefore trivially zero.

\subsection{Weak flow regime}
\begin{figure} %
   \centerline{ 
        \includegraphics[scale=0.45]{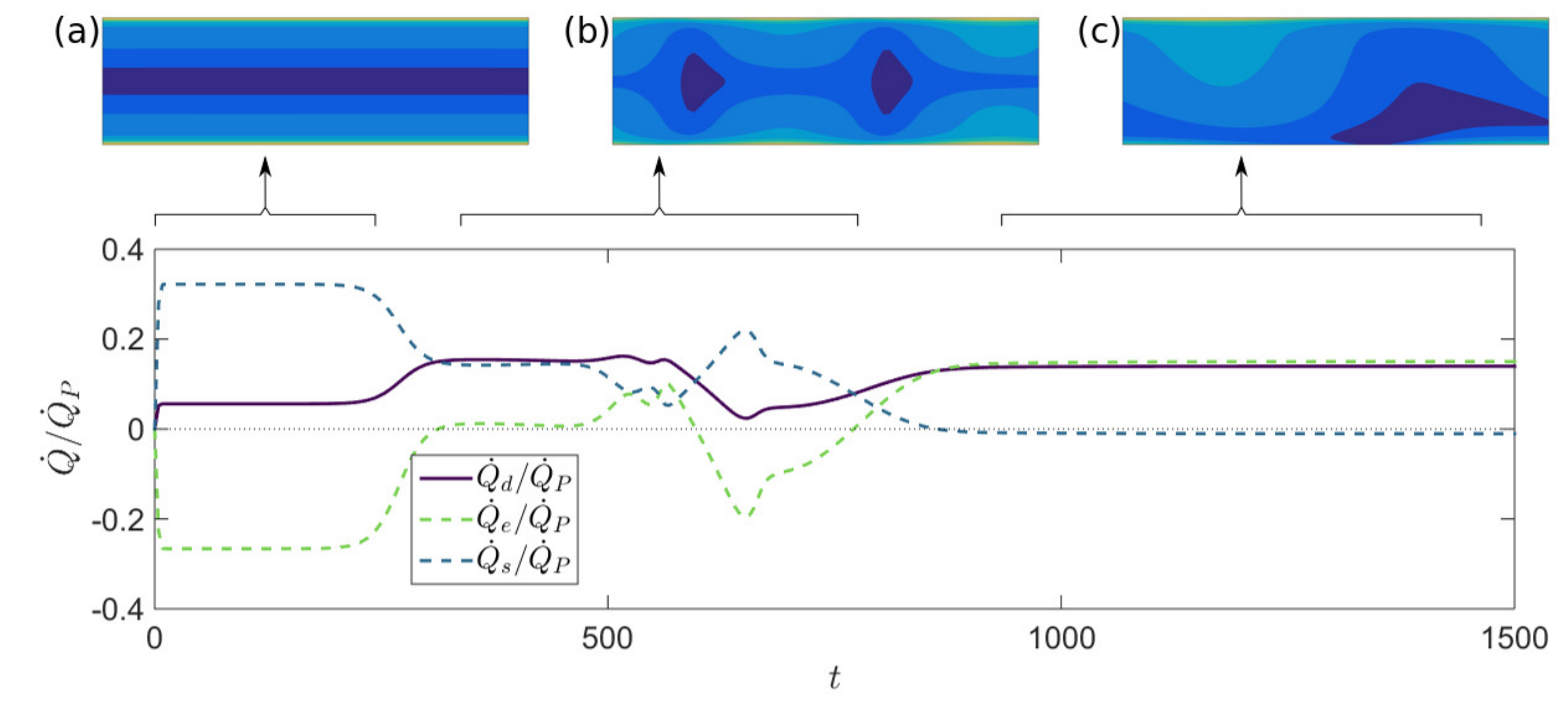}}
    \caption{Evolution of the particle-induced flow rate relative to the background flow rate, $\dot{Q}_d/\dot{Q}_P$ (solid line) and its self-induced (dashed blue) and externally-induced (dashed green) components, for the weak-flow regime ($\gamma_w=0.25$). Insets: Corresponding distribution of the particle density $\Phi$ in the three representative stages, (a)--(c).} \label{fig:Q_dots_low}
\end{figure}

We thus start by examining how the net flow rate is modified by the collective particle dynamics described in Sec.~\ref{sec:weak_flow} in the weak-flow regime ($\gamma_w=0.25$). 
Figure~\ref{fig:Q_dots_low} illustrates the temporal evolution of  the particle-induced flow rate $\dot{Q}_d$ and its two components, $\dot{Q}_s$ and $\dot{Q}_e$. Three successive plateaus are identified which correspond to the different phases of the evolution of the suspension, and are illustrated by the corresponding contour plots of the particle density.   

Shortly after the beginning of the simulation and up to $t\approx 230$ the solution is in the symmetric 1D fixed point (figure~\ref{fig:Q_dots_low}a). During this phase, $\dot{Q}_s/\dot{Q}_P>0$: the net flow rate is enhanced by the active self-induced particle stress. We note that the particles considered here are pushers ($\alpha_s<0$), and this result is therefore consistent with the reduction in effective viscosity at low shear rate identified theoretically \citep{Hatwalne2004,Saint_2010,Matilla2016} and experimentally \citep{Gachelin2013,Lopez2015} for suspensions of elongated pushers.

In contrast, $\dot{Q}_e/\dot{Q}_P<0$: the externally-induced stress tends to hinder the background flow, increasing the effective viscosity. Some observations to explain this result are in order. 
For chemotactic particles $\xi_r>0$ has two important consequences: (i) The sign of the externally-induced stresslet is positive, $\alpha_e>0$, as both quantities depend on the mobility contrast $M^-$, see Eqs.~\eqref{nondim_u0_xit_xir} and \eqref{alphas};  
(ii) particles tend to be positively aligned with the local chemical gradient, $\boldsymbol{G}$. 
Recalling the definition of $\mathbf{S}_e$,  
Eq.~\eqref{stresslet_dim}, the hydrodynamic signature of a particle directed along $\boldsymbol{p}$ and parallel to $\boldsymbol{G}$ is that of a puller swimmer oriented in the same direction. Clearly, the generated flow field will create a net flow in the opposite direction to the pusher-type self-induced flow ($\alpha_s<0$).  
Such tendency of puller-like swimmers to increase the effective viscosity is in agreement with previous experimental \citep{Rafai2010,McDonnell2015} and theoretical studies~\citep{Hatwalne2004,Matilla2016}.\\

At $t\approx 230$, the instability of the 1D symmetric state can clearly be seen (figure~\ref{fig:Q_dots_low}b) before the solution eventually reaches its 2D final state (figure~\ref{fig:Q_dots_low}c and figure~\ref{fig:Fi_C_lowSh}, bottom). 
Remarkably, during the transition from the 1D fixed point (a), passing through (b), up until the final state (c) the components of the active flow rate, $\dot{Q}_s$ and $\dot{Q}_e$, first decrease in magnitude and then change sign. This means that the chemically-induced polarisation and aggregation of the suspension (chemotactic instability) reverse the effect of each component of the active stress tensor on the effective viscosity: the pusher contribution now increases the viscosity ($\dot{Q}_s<0$) while the puller-like contribution of the chemically-induced component reduces it ($\dot{Q}_e>0$). To the best of our knowledge, this is the first prediction of a pusher and puller-like hydrodynamic signatures increasing and decreasing the effective viscosity, respectively, and results from the two-dimensional nature of the suspension organisation (in contrast with the purely one-dimensional settings considered usually).  
Note that the total particle-induced flow rate remains positive throughout the entire succession of dynamical phases outlined above, corresponding to a net reduction in effective viscosity in the weak-flow regime. This indicates the dominance of the chemically-induced stresses in the long-term 2D dynamics, which were subdominant during the initial phase and whose intensity depends on the local particle polarisation and chemical gradients.   

\begin{figure} 
   \centerline{
       \includegraphics[scale=0.32]{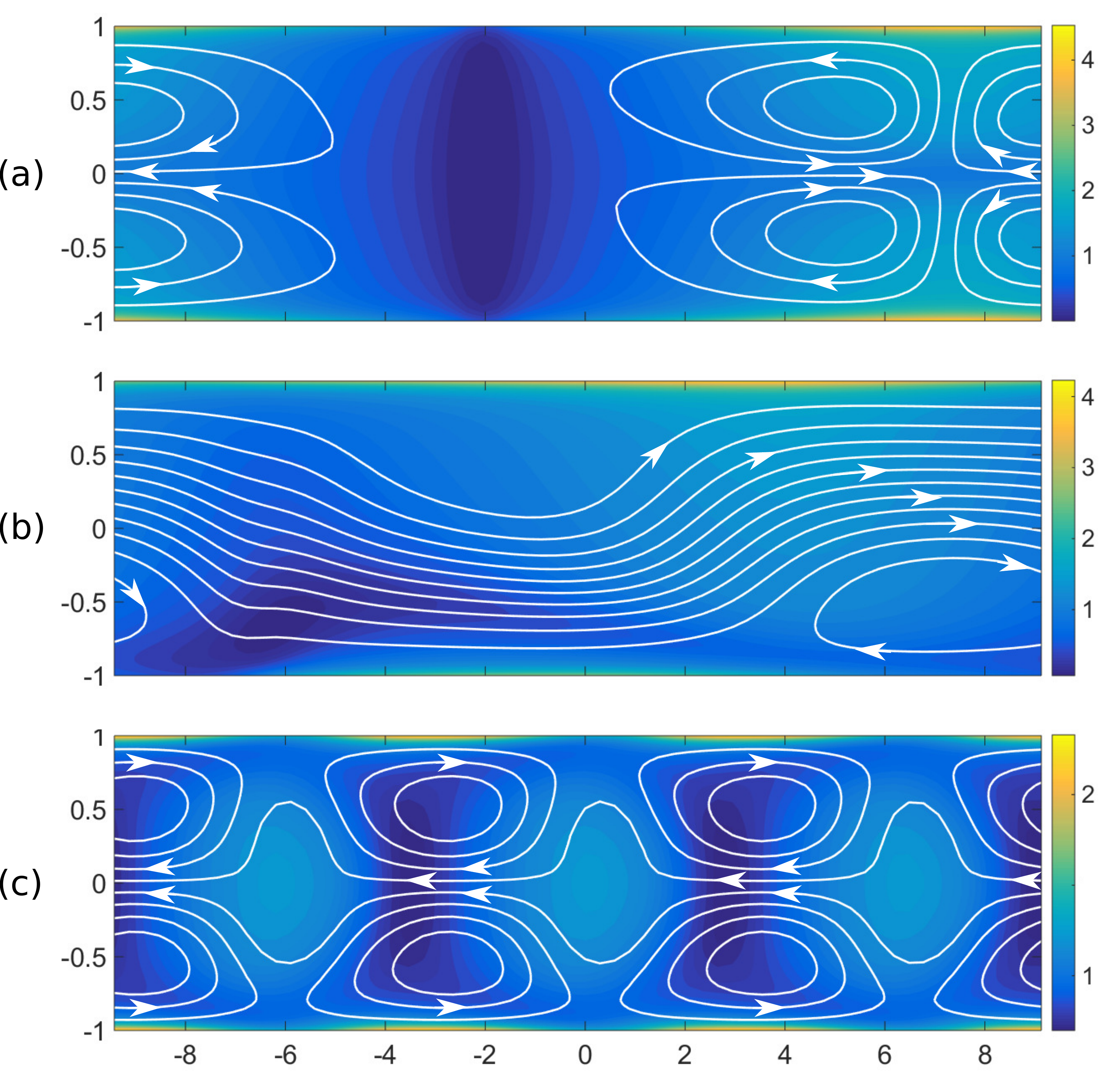}} 
    \caption{Streamlines of the disturbance flow generated by the particles, $\boldsymbol{u}_d$, for different background flow intensities, $\gamma_w$, at large times; the colorbar refers to the particle density $\Phi$ . (a) $\gamma_w=0$, $t=1300$; (b) $\gamma_w=0.25$, $t=1400$; (c) $\gamma_w=2.5$,  $t=600$ } \label{fig:flows} 
\end{figure}

The detailed particle-induced flow $\boldsymbol{u}_d$ observed at large times is reported on figure~\ref{fig:flows}. For all flow intensities, the presence of aggregates at the walls induces counter-rotating vortices at the two sides of the aggregates which are reminiscent of those found in other confined active systems~\citep{Wioland_2016,Shendruk_2017}. With no imposed flow (figure~\ref{fig:flows}a) all streamlines are closed: no net flow is induced, i.e., $\dot{Q}_d = 0$. Upon increasing $\gamma_w$ recirculation regions gradually disappear (figure~\ref{fig:flows}b), and open streamlines are found as the background flow breaks the left-right symmetry in the particles distribution and $\dot{Q}_d \neq 0$. 
\subsection{Moderate flow regime}
\begin{figure}
  \centerline{
        \includegraphics[scale=0.42]{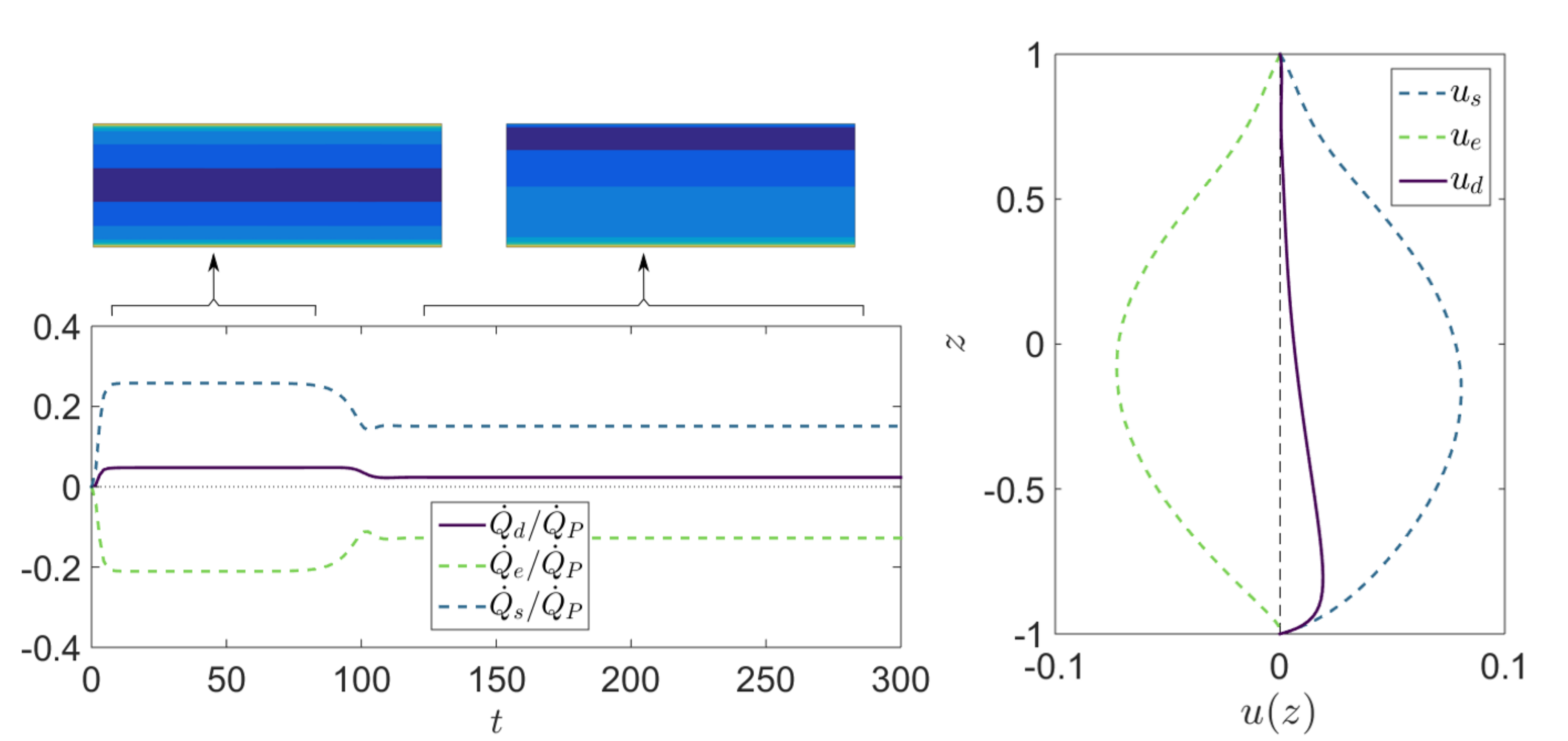} }
\caption{Left: Particle-induced flow rate in the moderate-flow regime ($\gamma_w=1$). Evolution in time of particle-induced flow rate relative to the imposed flow rate $\dot{Q}_P$ (solid line), and its self-induced ($\dot{Q}_s$) and externally induced ($\dot{Q}_i$) components (dashed lines). Representative particle density distributions are shown for the two main phases of the simulations (insets). Right: Particle-induced flow profiles of $u_s(z)$, $u_e(z)$ and $u_d(z)$ in the intermediate-flow regime ($\gamma_w=1$) at $t=300$.} \label{fig:u_profiles_int}
\end{figure}

In the moderate flow regime ($0.5\leq\gamma_w < 2.1$), the suspension's characteristics (e.g. particle and solute concentration) are uniform in the streamwise direction (figure~\ref{fig:Fi_C_asymm}) and so is the particle-induced flow, $\boldsymbol{u}_d=u_d(z)\boldsymbol{e}_y$. Its two components, $u_s$ and $u_e$, are observed to respectively enhance and hinder the background flow (figure~\ref{fig:u_profiles_int}), which is consistent with experimental observations and theoretical predictions for active suspensions in parallel flows and the sign of the associated stresslet intensities, $\alpha_s$ and $\alpha_e$. The effect on the suspension viscosity of pusher or puller swimmers does not depend on the symmetry property of the flow but are generic (i.e. they apply to both symmetric and asymmetric configurations). %

However, we note that $\dot{Q}_s$ and $\dot{Q}_e$ are both reduced in magnitude when the solution becomes asymmetric around $t=100$ (figure~\ref{fig:u_profiles_int}, left). In the symmetric state, the local chemical gradient is always pointing towards the nearest wall and its relative orientation with respect to the (anti-symmetric) background vorticity is the same throughout the channel. This does not hold any more as the top-down symmetry is broken: in part of the upper half of the channel, the chemical gradient is pointing downward, which reduces both $u_s$ and $u_i$ in that region, and moves their maxima in the region $z<0$ (figure~\ref{fig:u_profiles_int}, right).

Finally, unlike in the weak flow case, the self-induced component is now dominant and the overall effect is still to enhance the background flow and therefore to reduce the apparent viscosity of the active fluid, as can be seen from the positive sign of $\dot{Q}_d$ in figure~\ref{fig:u_profiles_int}. 
\subsection{Strong flow regime}
\begin{figure} 
     \centerline{
        \includegraphics[scale=0.4]{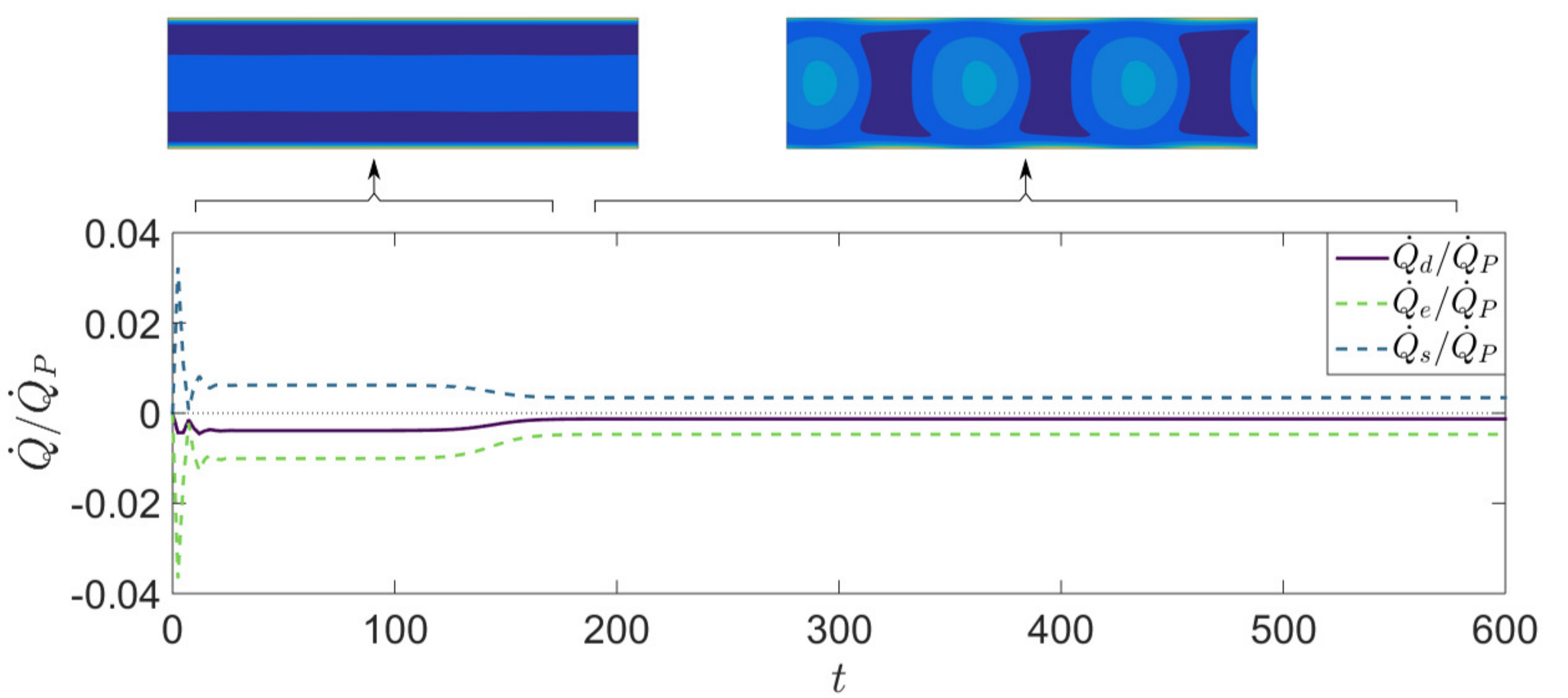}} 
        \caption{Particle-induced flow in the strong-flow regime ($\gamma_w=2.5$). Evolution in time of the particle-induced flow rate relative to the imposed flow rate $\dot{Q}_P$ (solid line), and its self-induced ($\dot{Q}_s$) and externally induced ($\dot{Q}_e$) components (dashed lines). Representative particle density distributions are shown for the two main phases of the simulations (inset).} \label{fig:Qdot_strong_flow}
\end{figure}

In the strong flow regime, the long-term solution is characterised by fast-moving aggregates. Similarly to the no-flow and weak flow cases, the formation of aggregates near the walls produces recirculating regions (figure~\ref{fig:flows}c, left). As the patterns emerge, a reduction of both $\dot{Q}_s$ and $\dot{Q}_e$ is visible in figure~\ref{fig:Qdot_strong_flow}, an effect that can be explained in total analogy with the weak flow case. However, the sign of neither of them is reversed due to the strong background vorticity, whose effects dominates and, qualitatively, the effect of the pusher and puller stresses on the total flow remain the same as in the streamwise-uniform configuration.

\subsection{Remarks on the effects of the disturbance flow}
\begin{figure} 
     \centerline{   
        \includegraphics[scale=0.48]{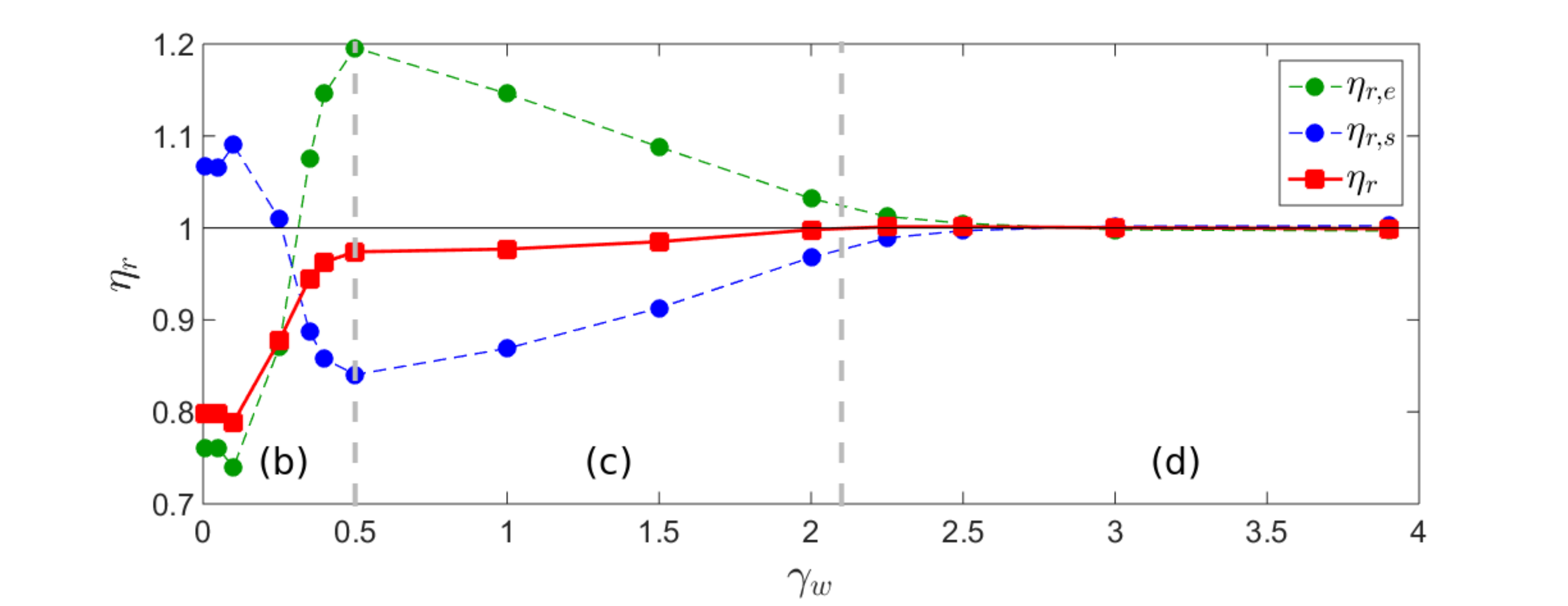}}
 \caption{Evolution of the viscosity $\eta_r$, Eq.~\eqref{eta_r_def} with the intensity of the background flow as measured at large times, when the solution is either time invariant, region (c), or a travelling wave, regions (b) and (d) (see Fig.~\ref{fig:ppt_scheme}).} \label{fig:eta_r_fy}
\end{figure}

As suggested by the numerical simulations (and confirmed in the subsequent analysis of the reduced model of Sec.~\ref{sec:ROM} where $\boldsymbol{u}_d$ is neglected), the effect of the particle-induced flow on the suspension's organisation is qualitatively negligible. 
However, $\boldsymbol{u}_d$ impacts significantly the long-term effective viscosity $\eta_r$, whose variations with $\gamma_w$ are displayed on figure~\ref{fig:eta_r_fy}, together with that obtained by accounting exclusively for self-induced stresses ($\eta_{r,s}$) or externally-induced ones ($\eta_{r,e}$). Note that these do not contribute additively to $\eta_r$, see Eq.~\eqref{eta_r_def}.

Above a certain flow strength (here $\gamma_w \approx 3$) $\eta_r\approx 1$, showing that the particle's hydrodynamic forcing is negligible and the background shear stress dominates. 
In this regime, the passive stress due to the inextensibility of the particles (which is not considered here) is also expected to be important and to increase the effective viscosity for pusher and puller suspensions regardless of the particles' shape \citep{Krieger_1959,Lopez2015,McDonnell2015,Rafai2010,Matilla2016}. 

We therefore focus our attention on the active contributions to the effective viscosity at lower flow strengths, namely in regions (b) and (c) of figure~\ref{fig:eta_r_fy}.  In region (c), the puller-like contribution to the viscosity, $\eta_{r,e}$, contributes positively to $\eta_r$ resulting into a decrease of $\eta_{r,e}$ with $\gamma_w$ (shear-thinning region), in line with previous studies on pullers \citep{Rafai2010,Matilla2016}. The trend is reversed in region (b), for weak flows. Here, starting at $\gamma_w=0.1$, $\eta_{r,e}$ increases with $\gamma_w$ resulting in a shear-thickening region. This behaviour is associated with 2D dynamics of the suspension and linked to the local appearance of a new preferential direction (other than the wall normal direction and the flow direction), represented by the chemical gradient. 
The same behaviour, in a specular way, occurs to the pusher contribution, $\eta_{r,s}$, which goes from promoting shear-thinning at weak flows, region (b), to promoting shear-thickening at intermediate flows, region (c). 

The overall contribution $\eta_r$ is always negative but the shape of the curve $\eta_r(\gamma_w)$ might qualitatively change depending on the relative magnitude of the two opposite contributions, $\eta_{r,s}$ and $\eta_{r,e}$. 
The sign of $\eta_{r,s}$ and $\eta_{r,e}$, instead, depends on the relative importance between chemical reorientation and rotation by the flow, which always varies smoothly from flow-dominated to chemically-dominated below some non-vanishing value of $\gamma_w$. 
For this reason the shape of the curves of $\eta_{r,s}$ and $\eta_{r,e}$ against $\gamma_w$ are expected to be robust, at least qualitatively, 
representing one of the main contributions of this work.  

Finally we stress  the generality of the present results, beyond the particular case of Janus phoretic particles: the sign reversal of the puller and pusher footprints on the viscosity is indeed relevant to all suspensions of microswimmers  prone to exhibit chemotactic dynamics even if characterised by only one hydrodynamic signature. In other words, the curves of  $\eta_{r,s}$ and $\eta_{r,e}$ in figure~\ref{fig:eta_r_fy} can be seen as independent (qualitative) prediction of $\eta_r$ for suspensions of autochemotactic pusher and puller microorganisms, respectively.

%

\section{Reduced-order model} \label{sec:ROM}

We noted earlier that understanding the suspension's organisation does not require accounting for interparticle hydrodynamic interactions. 
This was further confirmed by setting $\alpha_s$ and $\alpha_e$ artificially to zero in the full kinetic model (i.e., forcing $\boldsymbol{u}_d=\textbf{0}$): the solution was found to be quantitatively similar to that obtained with $\alpha_s,\alpha_i\neq 0$ for all values of $\gamma_w$. 
This observation is true for a sufficiently high degree of confinement ($F=1$), and ceases to hold for higher values of $F$, as confirmed by few preliminary results mentioned in section~\ref{sec:overview}.

In order to identify the fundamental underlying physical ingredients of the suspension's organisation, we propose here a reduced model of a phoretic suspension based on a two-moment expansion of the probability distribution $\Psi(\boldsymbol{x},\boldsymbol{p},t)$, effectively describing the system in terms of the local particle density $\Phi$ and polarisation $\boldsymbol{n}$.   

%
\subsection{Derivation of the two-moment system}
In the following, we approximate $\Psi(\boldsymbol{x},\boldsymbol{p},t)$ by its truncated moment expansion in $\boldsymbol{p}$ in terms of its zeroth and first orientational moments, $\Phi(\boldsymbol{x},t)$ and $\boldsymbol{n}(\boldsymbol{x},t)$, yielding 
\begin{align}
\Psi(z,\boldsymbol{p}) = \frac{1}{2\pi}\Phi + \frac{1}{\pi}\boldsymbol{p}\boldsymbol{\cdot}\boldsymbol{n}. \label{Psi_mom_1}
\end{align}

The evolution equations for the first two moment intensities, $\Phi$ and $\boldsymbol{n}=(n_y,n_z)$, are obtained by taking the first two orientational moments of the Smoluchowski equation \eqref{EvolEqPsi}. 
Understanding the dynamics discussed in Sec.~\ref{sec:NumSym} does not require taking into account the disturbance flow induced by the particles, which suggests that a reduced model can be derived considering only the effect of the background flow, thereby reducing the system's complexity significantly as the flow is now completely imposed $\boldsymbol{u}=(U_P(z),0)$. The evolution equations of $\Phi$ and $\boldsymbol{n}$ are then obtained as
\begin{eqnarray}
 \frac{\partial \Phi}{\partial t} + \boldsymbol{u}\boldsymbol{\cdot}\nabla\Phi & = &
  - \frac{u_0}{F} \nabla\boldsymbol{\cdot} \boldsymbol{n} - \xi_t \left( \nabla C \boldsymbol{\cdot} \nabla\Phi + \Phi\nabla^2C \right) + d_x\nabla^2\Phi  \label{mom0_1mod_yt}, \\
 \frac{\partial \boldsymbol{n}}{\partial t} + \boldsymbol{u}\boldsymbol{\cdot} \left(\nabla\boldsymbol{n}\boldsymbol{\cdot}\boldsymbol{e}_y\right) & = &
  - \frac{u_0}{2F}  \nabla\Phi - \xi_t \left[ \nabla C\boldsymbol{\cdot} (\nabla\boldsymbol{n})^{\textrm{T}} + \boldsymbol{n}\nabla^2C \right] 
  + \frac{\xi_r \Phi}{2\rho} \nabla C   \nonumber \\ 
  && \mbox{} + d_x \nabla\boldsymbol{\cdot} (\nabla\boldsymbol{n})^{\textrm{T}} - d_p\boldsymbol{n} 
  + z\frac{\gamma_w}{2}\boldsymbol{n}\boldsymbol{\cdot}\left( \boldsymbol{e}_z\boldsymbol{e}_y - \boldsymbol{e}_y\boldsymbol{e}_z \right), \label{n_eqs} 
\end{eqnarray} 
subject to the boundary conditions at $z=\pm 1$
\begin{align}
 \frac{u_0 n_z}{F}  = d_x \frac{\partial \Phi}{\partial z}, \hspace{1cm}
  \frac{\partial n_y}{\partial z} = 0 , \hspace{1cm}
  \frac{u_0\Phi}{2F} = d_x \frac{\partial n_z}{\partial z}\boldsymbol{\cdot} \hspace{1cm} \label{BC_1234}
\end{align}
Eqs.~\eqref{mom0_1mod_yt}-\eqref{BC_1234} together with Eq.~\eqref{C_eq_nondim} and \eqref{BC_C} for the chemical concentration form a closed system of partial differential equations, referred to in the following as the two-moment system or reduced model. 
Using Eq.~\eqref{Psi_mom_1} to derive the system~\eqref{mom0_1mod_yt}-\eqref{BC_1234} from Eq.~\eqref{EvolEqPsi} implies approximating the next (second) orientational moment as
\begin{eqnarray}
    \mathbf{D}(\boldsymbol{x},t) = \langle \boldsymbol{pp}-\frac{\boldsymbol{I}}{2} \rangle \approx \frac{\Phi\boldsymbol{I}}{2}.    \hspace{1cm}   \label{def_nDH}
\end{eqnarray}

Similar closure schemes based on a truncated expansion on the basis of spherical harmonics are commonly used for analysis or simulations of active suspensions \citep{Brotto2013,Baskaran_Marchetti_2009,ezhilan_saintillan_2015,Matilla2016,Saintillan2013,Saint2019_big_num}. The truncation point though is commonly at the third mode (second moment) or after, which is necessary to compute the self-induced active stress as $ \mathbf{S}_s \sim \mathbf{D}$. 
Moreover, $\mathbf{D}$ is also required to study the evolution of the orientational distribution of elongated particles through Jeffrey's equation. Including the second moment is therefore necessary, for example, to study the hydrodynamic instability emerging in suspensions of elongated pushers \citep{Saint2019_big_num} or the rheological properties of active suspensions under externally-imposed flow \citep{Saint_2010,Hatwalne2004}.

In contrast, for a chemotactic suspension of spherical autophoretic swimmers, the leading order interaction routes of the particles with each other and with the background flow are captured already at the polarisation level.
Indeed, in our dilute model the chemical field generated by the swimmers is a function of $\Phi$ only, see Eq.~\eqref{C_eq_nondim}. Similarly the translational drift induced by the chemical gradient, $\xi_t\nabla_x C$, and that induced by the flow field through Faxen's law do not depend on the particle's orientation and therefore influence the evolution of the particle density regardless of the local polarisation, see Eq.~\eqref{mom0_1mod_yt}.
The self-propulsion velocity, $u_0\boldsymbol{p}$, as well as the vorticity- and chemically-induced rotations, $\frac{1}{2}\boldsymbol{\omega}\times \boldsymbol{p}$ and $\frac{\xi_r}{\rho}( \boldsymbol{p} \times \nabla_x C )\times \boldsymbol{p}$, depend on the particle's orientation and therefore the first orientational moment needs to be included to capture their effects. In turn, these terms associated with the rotational motion of the particles, which naturally are present in the equations for $\boldsymbol{n}$, Eq.~\eqref{n_eqs}, enter the equation for the particle density $\Phi$ only in the presence of self-propulsion and with non-zero polarisation density $\boldsymbol{n}$ (see the first term on the right-hand side of Eq.~\eqref{mom0_1mod_yt}). To conclude, by using Eqs.~\eqref{Psi_mom_1} and \eqref{def_nDH} we can account for the leading order particle interactions discussed in Sec.~\ref{sec:NumSym} in a minimal way. 
%

%
\subsection{1D equilibria of the two-moment system}
\begin{figure} 
   \centerline{
        \includegraphics[scale=0.5]{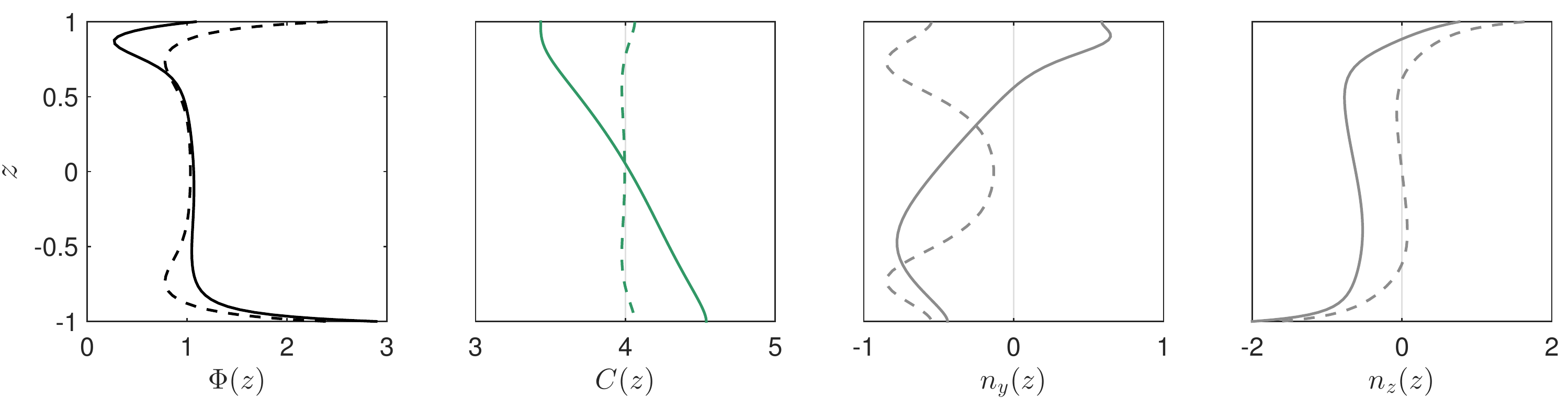}} 
    \caption{Symmetric and asymmetric 1D fixed points of the two-moment system for $\gamma_w=1.8$ for a chemotactic suspensions (the values of the other parameters match that of the full numerical simulations reported in the previous sections).} \label{fig:base_state}
\end{figure}

We first analyse the existence of steady 1D (i.e. $y$-invariant) solutions by marching in time the 1D version of the reduced model described above (i.e considering $\partial/\partial y=0$ for all the solution components) until a steady state is reached. Remarkably, we find that one symmetric and one asymmetric steady states coexist when the background flow is weak enough ($\gamma_w<1.9$), in agreement with the numerical simulations of the full model which predicted the coexistence of these states for  $\gamma_w<2.1$. 
Furthermore, the comparison of Figs.~\ref{fig:base_state} and \ref{fig:Fi_C_asymm}  shows that the asymmetric 1D fixed point of the reduced model retains the qualitative features of its counterpart in the full model, in particular the establishment of a nearly uniform wall-normal chemical gradient and particle polarisation across most of the channel width, that drive two competing effects, namely a chemotactic migration toward the chemically-rich wall and a phoretic drift toward the chemically-depleted one, the former effect being dominant. 

%
\subsection{Stability analysis of the two-moment system}
\begin{figure} 
  \centerline{                                       
        \includegraphics[scale=0.35]{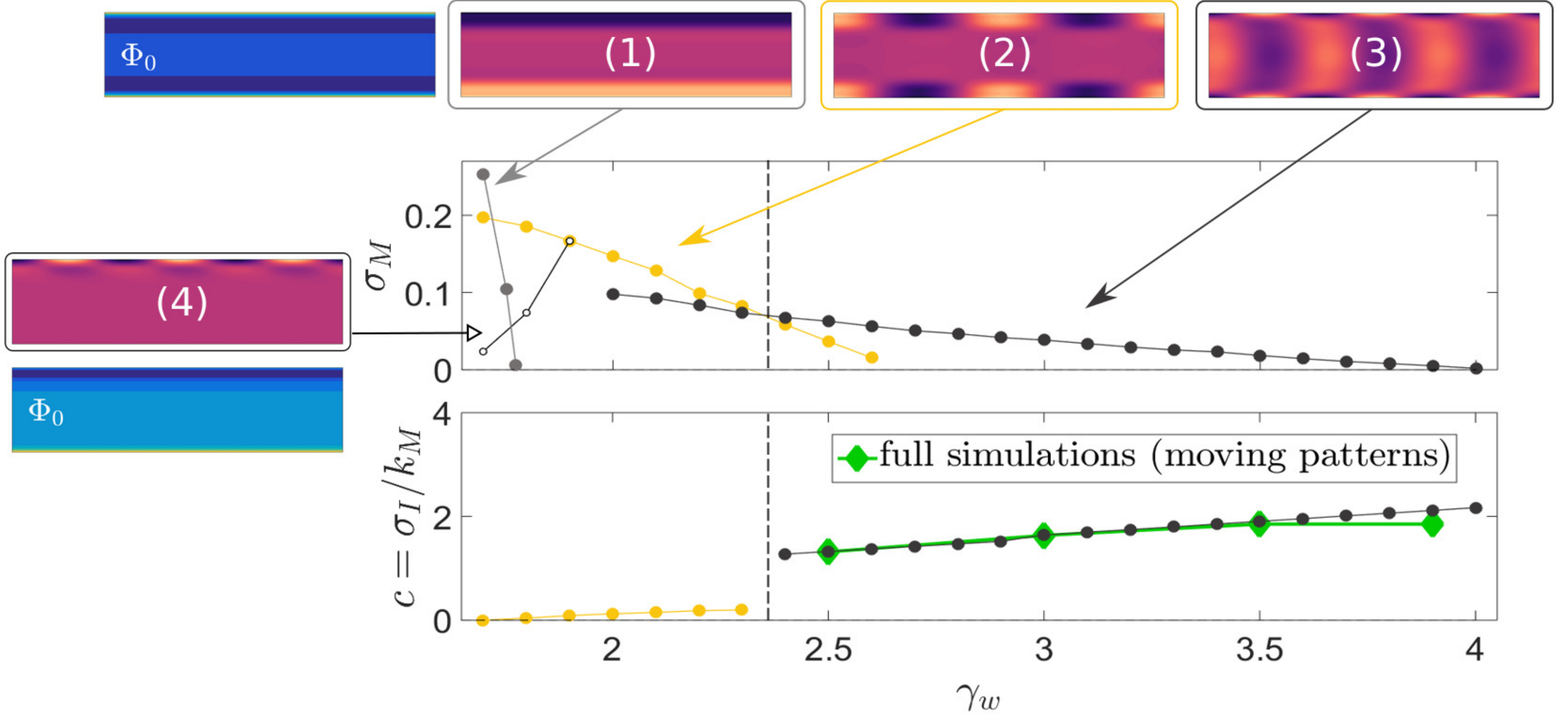}} 
\caption{Growth rate $\mathrm{Re}(\sigma)$ and wave speed $c=\mathrm{Im}(\sigma)/k$ of the most unstable modes as a function of flow rate. Open and filled symbols correspond respectively to  asymmetric and symmetric base states. Insets (1) to (4): Real part of the eigenmodes of the particle density, $\tilde{\Phi}$. The particle density of the corresponding 1D base states, $\Phi_0(z)$, is also plotted for reference. The vertical grey dashed line separates the regions where slow-moving wall patterns dominate (mode 2) in the reduced model, from the region where fast-moving patterns dominate (mode 3). Green diamonds: wave speed of \emph{fast} moving patterns measured in numerical simulation of the full Fokker-Planck equation.  } \label{fig:stab_an_chem}
\end{figure}

In this section we analyse the stability of this 1D fixed point with respect to streamwise perturbations:
\begin{align}
  \Phi = \Phi_0(z) + \epsilon \Phi'(y,z,t), \quad 
  \boldsymbol{n}=\boldsymbol{n}_0(z) + \epsilon\boldsymbol{n}'(y,z,t), \quad
  C = C_0(z) + \epsilon C'(y,z,t). \label{linearization}  
\end{align} 
After linearizing the two-moment system in the limit of small $\epsilon$, we seek solutions of the form $\Phi'(y,z,t) = \tilde{\Phi}(z,k)e^{(iky + \sigma t)}$ and analogously for $\boldsymbol{n}'$ and $C'$.
We denote by $\tilde{\boldsymbol{x}}$ the column state vector containing the $z$-discretised state variables $(\Phi',n'_y, n'_z, C')$ on a Gauss-Lobatto grid with $N+1$ points, 
\begin{align}
\tilde{\boldsymbol{x}} = [\tilde{\Phi}^{(1)}, \dots , \tilde{\Phi}^{(N+1)}, \tilde{n}_y^{(1)},\dots ,\tilde{n}_y^{(N+1)} , \tilde{n}_z^{(1)},\dots ,\tilde{n}_z^{(N+1)} ,\tilde{C}^{(1)},\dots ,\tilde{C}^{(N+1)}  ]^{\textrm{T}}.
\end{align} 
The linear stability of the model thus finally reduces to an eigenvalue problem of the form
\begin{eqnarray}
\sigma \tilde{\boldsymbol{x}} = \mathcal{L}\tilde{\boldsymbol{x}}, \label{Eig_pr}
\end{eqnarray}  
with $\mathcal{L}$ the linear operator accounting for the discretised linear reduced model modified adequately to account for the boundary conditions. This system is then solved numerically using a Matlab's algorithm based on the principle of minimized iterations~\cite{Arnoldi1951}.  
Figure~\ref{fig:stab_an_chem} reports the growth rate $\sigma_M=\textrm{Re}(\sigma)$ of the most unstable mode as a function of the strength of the imposed flow, measured by $\gamma_w$. We emphasize that the base state considered is different for each value of $\gamma_w$, and for $\gamma_w<1.9$, different symbols are used for the modes related to symmetric and asymmetric base states.  

In agreement with the full numerical simulations, the stability analysis shows the presence of unstable modes until $\gamma_w \approx 4$, above which the $y$-uniform solution is stabilised under the effect of strong background flow. For $1.7 < \gamma_w<2.4$, the most unstable mode associated with the symmetric base state is characterised by denser regions near the walls (see figure~\ref{fig:stab_an_chem}, inset (2)), which resemble the unstable modes observed during the transient dynamics of the full simulations for $\gamma_w<1.5$ (shown only for $\gamma_w=0$ in figure~\ref{fig:Fi_C_NoFlow}, top). The dominant mode for $\gamma_w\gtrsim 2.4$ (figure~\ref{fig:stab_an_chem}, inset (3)) closely resembles the moving patterns described in figure~\ref{fig:Fi_C_highSh} (bottom).

The eigenvalues associated with these two families of modes have a nonzero imaginary part $\textrm{Im}(\sigma)=\sigma_I$ that grows linearly with $\gamma_w$, which is consistent with a wave solution travelling in the streamwise direction with speed $c$ proportional to the background flow. More precisely, $c$ can be estimated from the stability analysis of the reduced model as  
\begin{eqnarray}
  c = \left(\frac{\sigma_I}{2\pi}\right)  \left(\frac{2\pi}{k_M}\right),
\end{eqnarray}  
where $2\pi/k_M$ is the wavelength of the most unstable mode. This estimate of $c$ is  in agreement with the actual wave speed measured in full simulations in the strong flow regime, i.e., \emph{fast} moving patterns equivalent to modes (3) in the reduced model, as plotted in figure~\ref{fig:stab_an_chem} (bottom).
The wave speed associated with modes (2), similar to the wall patterns observed in the full numerical simulation, is significantly smaller than that of  modes (3) which is also consistent with the full simulations.

Focusing on the range of flow rates $1.7<\gamma_w<1.9$ in figure~\ref{fig:stab_an_chem} (top), two observations can be made. First, an unstable branch of asymmetric modes uniform in $y$ (i.e. 1D modes) is present at low flow rates, modes (1), and is more unstable at low flow rates than the wall patterns, modes (2). Branch (1) assesses the stability of the symmetric 1D fixed point to 1D perturbations and is consistent with the observations of the full numerical simulations (see figure~\ref{fig:ppt_scheme}): 
it confirms the existence of an intrinsic instability of symmetric configurations for sufficiently weak flows, as reported in Secs.~\ref{sec:weak_flow} and \ref{sec:AsymState}. 

Also, a branch of unstable 2D modes relative to the asymmetric base states (open circles) tend towards negative growth rates as $\gamma_w$ decreases, suggesting the existence of a stable region of the asymmetric base state, qualitatively corresponding to the intermediate flow regime, line (c) in figure~\ref{fig:ppt_scheme}.   

Finally, for background flows weaker than $\gamma_w<1.7$, solving the two-moment system leads to solutions with negative particle density, and therefore loses physical meaning. This is not surprising as the reduced model, which is based on a truncated two-mode expansion of the full distribution function, loses the properties built-in the structure of the original conservation Eq.~\eqref{EvolEqPsi}. 

The reduced model and its stability analysis are therefore in qualitative agreement with the complete model, indicating that the diverse collective dynamics observed can be rationalised without accounting for hydrodynamic interactions between particles, and using a very simplified representation of the probability distribution, in particular regarding the orientation characteristics, simply described here by the local polarisation vector $\boldsymbol{n}(\boldsymbol{x},t)$. This is quite different from the representation of collective dynamics in bulk quiescent flows~\citep{Traverso2020}, suggesting that a transition towards bulk-like dynamics will take place by increasing the ratio between the channel width and the intrinsic length scale of the suspension, $F$, which is confirmed by unreported preliminary results and shall be one aspect of future investigations. 
\section{Conclusions} \label{sec:conclusions}

This work proposes an in-depth investigation into the self-organization dynamics of a dilute autochemotactic phoretic suspension under the combined effects of confinement inside a channel and of a pressure-driven shear flow. It also offers an analysis of the effect of the particles' hydrodynamic forcing on the resulting total flow rate for fixed pressure forcing, i.e. a measure of the effective viscosity of the active fluid. As the background flow intensity is increased, our results show a smooth transition from a weak flow regime, where the particles self-organise into spatio-temporal patterns, to a flow-dominated regime, where chemical interactions between particles are subdominant with respect to the transport and rotation by the imposed background flow that drives their individual and collective dynamics. 
Starting from an initial perturbation of the isotropic suspension configuration,  confined autophoretic suspensions quickly organize across the channel width leading to the emergence of 1D (i.e., stream-wise uniform) equilibria distributions associated with a top-down (i.e., cross-stream) symmetry-breaking. When these 1D configurations are unstable to streamwise perturbations due to chemical interactions among particles, the final long-term particle organization is inhomogeneous along the axis of the channel, and characterised by either wall aggregates or fast moving patterns, depending on the intensity of the background flow. 

The impact of the suspension's dynamics on the fluid's effective rheology (defined as the modification of the total flow rate for a fixed pressure forcing) was further examined in two distinct ways.   
The first one focused on the fluid's response to quasi-static variations of the external forcing (i.e. over time scales much larger than that of the intrinsic dynamics, see figure~\ref{fig:eta_r_fy}). Our results show that the collective organisation of the autochemotactic phoretic suspensions considered here tend to reduce its effective viscosity, particularly in the regime of lower flow forcing. Yet, the detailed analysis of the self-induced and externally-induced contributions to the particle-induced flow, shows a complex evolution of these two parts, respectively associated with the particle self-propulsion and its response to the chemical gradients generated by other particles. They are observed to contribute to the total flow rate in opposite direction to each other, but also reverse the sign of their individual contribution as the flow rate is gradually increased, as a result of the fundamentally-different organization of the particles in each regime. These observations demonstrate how inter-particle interactions and the consequent collective dynamics strongly influence the effective rheology of active fluids, which is an aspect vastly unexplored in the literature so far, with many potential engineering applications. 

For a given flow intensity, we also analysed the rheology of active suspensions by characterising the evolution of the particle-induced flow rate as the collective organisation of the particles evolve in time (see figures~\ref{fig:Q_dots_low}, \ref{fig:u_profiles_int} and \ref{fig:Qdot_strong_flow}). Our results suggest the possibility to control in real time the viscosity of an active fluid via the particles' collective dynamics. These can indeed be controlled, for example, by means of external chemical \citep{hong2007chemotaxis,palacci2010_PRL,palacci2013living} or optical \citep{sen2009chemo,Rafai2016_rapidPRE} signals, thus paving the way for new microfluidic applications involving either synthetic or biological microswimmers. 

 The particular configuration chosen for our study (i.e. a pressure-driven channel shear flow) exposes the particles to a non-uniform shear, and as a result, our characterisation of the fluid's rheology is phenomenological and macroscopic, although motivated by many technological applications which may be interested in reducing (or increasing) the flow rate in a thin channel for fixed pressure drop. Several of the self-organisation features identified here are directly linked to the non-uniformity of the shear and may not hold in other settings. An alternative approach to the characterisation of the suspension's rheology consists in analysing its response to a uniform shear rate (as in most rheometers).
 
 The shape of the particles substantially affects their response to the hydrodynamic field; as a result, a potential extension of the present work would consider suspensions of anisotropic (e.g. rod-like) particles  for which hydrodynamic reorientation in shear plays a significant in the organisation of the suspension and the particle-induced flow.
 
 We finally remark that our results are obtained in the dilute limit where far-field interactions are dominant. However, the control of the fluid's properties for technological applications would likely require to maximize the hydrodynamic forcing by the particles, which might naturally call for denser suspensions. In this case, the near-field steric, chemical and hydrodynamic interactions (among particles and with the boundaries) will also play an important role, and particle-based approaches~\citep[e.g.,][]{uspal2015,lambert2013active} may become more relevant despite their substantially higher computational cost.

\section*{Acknowledgements}
The authors thank L. Lesshafft for helpful conversations on the linear stability analysis of the reduced order model, and F. Picella on the treatment of nonlinear boundary conditions. This work was supported by the European Research Council (ERC) under the European Union Horizon 2020
research and innovation program (Grant Agreement No. 714027 to S.M.).

\appendix

\section{Single particle's dynamics} \label{app:A}
This appendix summarizes the characteristics of the individual particle's motion and stresslet (i.e., its dominant far-field hydrodynamic signature) in response to its own activity (Sec.~\ref{app:A1}) or an externally-imposed chemical gradient (Sec.~\ref{app:A2}).

 For an isolated force- and torque-free particle in an unbounded flow, the translational and rotational velocities of the colloid $\boldsymbol{U}$ and $\boldsymbol{\Omega}$, and  particle stresslet, $\hat{\mathbf{S}}$, are obtained from the surface slip velocity of the colloid, $\boldsymbol{u}^*$, as ~\citep{Stone1996,Lauga2016}
\begin{eqnarray}
\boldsymbol{U}=-\langle\boldsymbol{u}^*\rangle_{\partial V}, \quad \boldsymbol\Omega=-\frac{3}{2R}\langle \hat{\boldsymbol{r}}\times\boldsymbol{u}^*\rangle_{\partial V}, \quad
\hat{\mathbf{S}}=-10\eta\pi R^2 \langle  \hat{\boldsymbol{r}}\boldsymbol{u}^* + \boldsymbol{u}^* \hat{\boldsymbol{r}} \rangle_{\partial V}   \label{Ind_vel}, 
\end{eqnarray}
where $\langle  \rangle_{\partial V}$ indicates averages over the surface of the particle, $\partial V$, and $\hat{\boldsymbol{r}}=\boldsymbol{r}/r$ is the normal unit vector used to parametrize the particle's surface.  

The phoretic slip velocity at the particle's surface is obtained from the local chemical field and mobility distribution on the surface of the colloid \citep{Golestanian2007,Michelin2014}, $C(\hat{\boldsymbol{r}})$ and $M(\hat{\boldsymbol{r}})$, respectively, as
\begin{eqnarray}
   \boldsymbol{u}^* = M( \hat{\boldsymbol{r}})(\boldsymbol{I}-\hat{\boldsymbol{r}} \hat{\boldsymbol{r}})  \boldsymbol{\cdot}   \nabla_x C|_{r = R} \label{u_slip}.
\end{eqnarray}

\subsection{Self-propulsion and self-induced stresslet}\label{app:A1} 
The activity distribution is piecewise uniform with 
\begin{eqnarray}
    A(\hat{\boldsymbol{r}}) = A_b \ \ \ \textrm{for} \ \ \ \hat{\boldsymbol{r}}\boldsymbol{\cdot}\boldsymbol{p}<0, \ \ \ \textrm{and} \ \ \ 
    A(\hat{\boldsymbol{r}}) = A_f \ \ \ \textrm{for} \ \ \ \hat{\boldsymbol{r}}\boldsymbol{\cdot}\boldsymbol{p}>0 ,
\end{eqnarray}
and similarly for the mobility distribution $M(\hat{\boldsymbol{r}})$, where $\boldsymbol{p}$ is the unit vector pointing toward the front of the particle and along its axis of symmetry. As in the main text, we define $A^\pm = A_b \pm A_f$ the total activity and activity contrast, respectively, and similar definitions for the mobility equivalents, $M^\pm$.

At the particle scale, $r=O(R)$, convective transport of solute is subdominant, so that the chemical field outside the particle due to its own chemical activity $A(\hat{\boldsymbol{r}})$ is the solution of a diffusion (Laplace) problem~\citep{Traverso2020}  
\begin{equation}
   D_c \nabla^2_x C=0, \qquad \textrm{with}\quad D_c  \hat{\boldsymbol{r}} \boldsymbol{\cdot} \nabla_xC|_{r=R}=- A(\hat{\boldsymbol{r}}) \ \ \ \textrm{and} \ \ \ C|_{r\rightarrow \infty} = 0,
\end{equation}
whose solution is obtained for the particles considered here, as~\citep{Golestanian2007} 
\begin{eqnarray}
  C = \left( \frac{2\pi R^2 A^+}{D_c} \right)\frac{1}{4\pi r} - \left( \frac{3\pi R^3 A^-}{2 D_c} \right) \left( \frac{\boldsymbol{p}  \boldsymbol{\cdot} \boldsymbol{r}}{4\pi r^3} \right) + \sum_{m=2}^{\infty} \frac{A_mR}{(m+1)D_c} \left(\frac{R}{r}\right)^{m+1} P_m(\boldsymbol{p}  \boldsymbol{\cdot} \hat{\boldsymbol{r}}) ,  \label{C_exp} 
\end{eqnarray}
with $P_m$ the Legendre polynomial of degree $m$. Coefficients $A_m$ are obtained as projections of the activity distribution, i.e. $A_m= (m+1/2) \int_{-1}^{1} A(\hat{\mu})P_m(\hat{\mu})\textrm{d}\hat{\mu}$, and for hemispheric swimmers $A_m=0$ for even $m$.
Substitution of Eq.~\eqref{C_exp} into Eqs.~\eqref{u_slip} and \eqref{Ind_vel} provides
\begin{eqnarray}
   \boldsymbol{U} = U_0\boldsymbol{p}\quad\textrm{with}\quad U_0 = \frac{A^-M^+}{8D_c}, \label{SelfP_dim}
\end{eqnarray}
as well as the self-induced stresslet
\begin{align}
\hat{\mathbf{S}}_s&=\hat{\alpha}_s \left(\boldsymbol{pp}-\frac{\boldsymbol{I}}{3}\right), \ \ \ \ \textrm{with} \ \ \ \  
        \hat{\alpha}_s = - \frac{10\pi \eta a^2 \kappa M^- A^-}{D_c} \hspace{3mm}
        \label{sigma_s}\\
        \textrm{and    }&\kappa = \frac{3}{4}\sum_{m=1}^{\infty} \frac{2m+1}{m+1} \left[ \int_0^1P_m \textrm{d}\hat{\mu} \right] \left[ \int_0^1 \hat{\mu} (1-\hat{\mu}^2)P'_m \textrm{d}\hat{\mu} \right]\approx 0.0872.
  \end{align}
Finally, due to the problem's symmetry there is no self-induced rotation, $\boldsymbol{\Omega}_s=0$.

\subsection{Externally induced drifts and stresslet} \label{app:A2}

We now consider the slip velocity and resulting drifts and stresslets, generated on a particle with no surface activity, by an external concentration $ C_{e} \sim C_{\infty} + \boldsymbol{G}\boldsymbol{\cdot}\boldsymbol{r}$ surrounding the particle. 
Because of the presence of the particle, the external field is modified in its vicinity as
\begin{eqnarray}
     C_e = C_{\infty} + \boldsymbol{G}\boldsymbol{\cdot}\boldsymbol{r}  \left( 1+\frac{R^3}{2 r^3} \right). \label{ext_C_1}
\end{eqnarray}
Substitution of this result into Eq.~\eqref{u_slip}, provides the externally-imposed slip velocity, $\boldsymbol{u}^*_e$, and substitution into Eq.~\eqref{Ind_vel} provides the translational and rotational drifts experienced by the force- and torque-free particle, respectively,  as
\begin{align}
\boldsymbol{U}_e&=\chi_t\boldsymbol{G},\qquad\textrm{with   } \quad\chi_t = -\frac{M^+}{2},\\
\boldsymbol{\Omega}_e&=\chi_r\boldsymbol{p}\times\boldsymbol{G}, \,\textrm{with }\quad \chi_r=\frac{9}{16}\frac{M^-}{R}\cdot
\end{align}
Note that the translational drift $\boldsymbol{U}_e$ does not include any component along $(\boldsymbol{G}\cdot\boldsymbol{p})\boldsymbol{p}$ in contrast with more generic particles~\citep{Kanso2019}, since the second Legendre projection $M_2$ of the mobility distribution vanishes for hemispherically-coated particles.

Similarly, the stresslet component due to the external gradient is obtained from Eq.~\eqref{Ind_vel} as
\begin{eqnarray}
    \hat{\mathbf{S}}_e = -\frac{15\eta}{4}  \int_{\partial V} M( \hat{\boldsymbol{r}}) \Big\{ \left[ (\boldsymbol{I}- \hat{\boldsymbol{r}} \hat{\boldsymbol{r}})  \boldsymbol{\cdot} \boldsymbol{G} \right]  \hat{\boldsymbol{r}} +  
    \hat{\boldsymbol{r}} \left[ (\boldsymbol{I}- \hat{\boldsymbol{r}} \hat{\boldsymbol{r}})  \boldsymbol{\cdot} \boldsymbol{G} \right] \Big\} \textrm{d}A \label{deriv_Sind_1}.
\end{eqnarray}
For piecewise uniform mobility distribution on each hemisphere, the integral in Eq.~\eqref{deriv_Sind_1} can be conveniently rewritten as the sum of two contribution, on the front ($\hat{\boldsymbol{r}} \boldsymbol{\cdot}\boldsymbol{p}>0$) and back hemispheres ($\hat{\boldsymbol{r}} \boldsymbol{\cdot}\boldsymbol{p}<0$). Using
\begin{eqnarray}
\int_{\hat{\boldsymbol{r}}\boldsymbol{\cdot}\boldsymbol{p}>0}\hat{\boldsymbol{r}}\,\mathrm{d}A=\pi R^2\boldsymbol{p},\qquad \int_{\hat{\boldsymbol{r}}\boldsymbol{\cdot}\boldsymbol{p}>0}\hat{\boldsymbol{r}}\hat{\boldsymbol{r}}\hat{\boldsymbol{r}}\,\mathrm{d}A = \frac{\pi R^2\boldsymbol{p}}{4}\left[\boldsymbol{p}\boldsymbol{I}+\boldsymbol{I}\boldsymbol{p}+(\boldsymbol{I}\boldsymbol{p})^{\textrm{T}_{23}}\right]
\end{eqnarray}
where $\mathbf{A}^{\textrm{T}_{23}}$ is the transpose of the third-order tensor $\mathbf{A}$ with respect to its last two indices (the integrals on $\hat{\boldsymbol{r}}\boldsymbol{\cdot}\boldsymbol{p}<0$ are obtained by changing $\boldsymbol{p}$ into $-\boldsymbol{p}$), the induced stresslet is finally obtained as
\begin{eqnarray}
        \hat{\mathbf{S}}_e   = \hat{\alpha}_e \left[ \boldsymbol{Gp} + \boldsymbol{pG} + (\boldsymbol{G}  \boldsymbol{\cdot} \boldsymbol{p}) (\boldsymbol{pp}-\boldsymbol{I}) \right],\qquad \textrm{with} \qquad  \hat{\alpha}_e = \frac{15}{8}\eta R^2 \pi M^-\label{Ind_stresslet2}   .
\end{eqnarray}

\end{document}